\documentclass[10pt,amsmath,amssymb,nofootinbib,twoside,twocolumn,superscriptaddress,floats,floatfix,aps,prd,preprintnumbers]{revtex4-2}
\usepackage[british]{babel}
\usepackage{xcolor}
\usepackage{booktabs}
\usepackage{latexsym}
\usepackage{amssymb}
\usepackage{amsmath}
\usepackage{graphicx}
\usepackage{tabularx}
\usepackage{cancel}
\usepackage{hyperref}

\hypersetup{
    colorlinks=true,
    linkcolor=red,
    citecolor=blue,
    filecolor=black,
    urlcolor=blue
}
\usepackage[capitalise]{cleveref}
\usepackage[utf8]{inputenc}
\usepackage[normalem]{ulem}
\def\beq#1\eeq{\begin{align}#1\end{align}}

\newcommand{\mms}{M_{\mbox{\tiny MS}}} 

\newcommand{\dd}{\text{d}}
\newcommand{\ee}{\text{e}}
\newcommand{\ii}{\text{i}}

\newcommand{\MP}[1]{{\color{violet}#1}}

\definecolor{darkgreen}{RGB}{0,170,0}

\begin{document}

\newcommand{\UniCa}{\affiliation{Dipartimento di Fisica, Universit\`a di Cagliari, Cittadella Universitaria, 09042 Monserrato, Italy}}
\newcommand{\INFNCa}{\affiliation{INFN, Sezione di Cagliari, Cittadella Universitaria, 09042 Monserrato, Italy}}
\newcommand{\INFNROMA}{\affiliation{INFN, Sezione di Roma, Piazzale Aldo Moro 2, 00185, Roma, Italy}}
\newcommand{\Sapienza}{\affiliation{Dipartimento di Fisica, Sapienza Università 
	di Roma, Piazzale Aldo Moro 5, 00185, Roma, Italy}}
\newcommand{\inafoac}{\affiliation{INAF - Osservatorio Astronomico di Cagliari, Via della Scienza 5, 09047 Selargius (CA), Italy}}
\newcommand{\gssi}{\affiliation{
Gran Sasso Science Institute (GSSI), Viale F. Crispi 7, L'Aquila (AQ), I-67100, Italy}}
\newcommand{\infngssi}{\affiliation{INFN - Laboratori Nazionali del Gran Sasso (LNGS), L'Aquila (AQ), I-67100, Italy}}

\author{Mariano~Cadoni}
\email{mariano.cadoni@ca.infn.it}
\UniCa\INFNCa

\author{Andrea~Cozzumbo}
\email{andrea.cozzumbo@gssi.it}
\gssi\infngssi

\author{Riccardo~Murgia}
\email{riccardo.murgia89@unica.it}
\UniCa\INFNCa\inafoac

\author{Matteo~Pili}
\email{matteo.pili@inaf.it}
\inafoac\UniCa\INFNCa

\author{Mirko~Pitzalis}
\email{mirko.pitzalis@ca.infn.it}
\UniCa\INFNCa

\author{Andrea~P.~Sanna}
\email{asanna@roma1.infn.it}
\Sapienza\INFNROMA

\author{Matteo~Schulz}
\email{matteo.schulz@gssi.it}
\gssi\infngssi

\author{Ilaria~Usai}
\email{ilaria.usai@ca.infn.it}
\UniCa\INFNCa

\title{Anisotropic Fluid Inhomogeneous Cosmology}
\begin{abstract}
 We propose a new framework for inhomogeneous cosmology based on General Relativity sourced by an anisotropic fluid. The model is parametrized by a single input function, the radial Misner--Sharp mass distribution at different redshifts, and admits exact solutions to Einstein's equations. Going beyond the standard assumption of small metric perturbations, it provides a fully relativistic framework to describe the non-linear regime of cosmic structure formation. The Friedmann-Lema\^itre-Robertson-Walker (FLRW) cosmology is recovered both at early times and -- on large scales only -- at late times. Our Anisotropic Fluid Inhomogeneous Cosmology (AFIC) model provides a natural mechanism through which matter inhomogeneities give rise to late-time acceleration. The extensive term driving this effect does not enter the exact metric solution and therefore does not contribute to the local gravitational potential. Rather, it emerges as a genuinely large-scale effect induced by the inhomogeneous distribution of matter at late times. This mechanism may provide a physical interpretation of the results from the Dark Energy Spectroscopic Instrument (DESI). Finally, we present a simple analytical realization of the model and use it to provide an estimate of the present-day effective dark energy density.
\end{abstract}

\maketitle

\section{Introduction}

Current cosmological probes have become precise enough to test foundational assumptions of the standard $\Lambda$ cold dark matter (CDM) model, such as the interpretation of dark energy (DE) as a cosmological constant $\Lambda$~\cite{DESIDR2II,DESI2024VI,CosmoVerseNetwork:2025alb,CANTATA:2021asi,Brax:2017idh,Copeland:2006wr,Linder:2005in,Peebles:2002gy,Wu:1998ad,Carroll:1991mt}, and the statistical homogeneity and isotropy of the universe~\cite{Blomqvist:2008ud,shafieloo_clarkson_2010,Colin:2010ds,Scrimgeour:2012wt,Alonso:2014xca,Camarena:2021mjr,Aluri:2022hzs,Dam:2022wwh,Secrest:2022uvx,Tang:2023kzs,Bengaly:2024ree,Hu:2024qnx,Seifert:2024bqr,Courtois:2025xcs,Heinesen:2026wwy,Koksbang:2026jxb,Koksbang:2026wvh}. In fact, although at sufficiently large distances our universe appears homogeneous and isotropic, there is growing evidence of inhomogeneities and anisotropies in the large-scale structure (LSS) of matter distribution~\cite{Hogg:2004vw,Yadav:2005vv,Maartens:2011yx,2013MNRAS.429.2910C,Perivolaropoulos:2014lua,Pandey:2015xea,Bengaly:2016amk,Bengaly:2017slg,Bengaly:2018ykb,Secrest:2020has,Krishnan:2021jmh,Dias:2023qrl,Lopez:2024kih,Sawala:2025ajg,Labini:2025dnc}.
This complexity is further highlighted by persistent tensions in the amplitude of the late Integrated Sachs-Wolfe (ISW) effect, with cross-correlations between Cosmic Microwave Background (CMB) temperature maps and LSS tracers pointing to unexpectedly high contributions from large-scale anisotropies~\cite{Granett:2008ju,Kovacs:2018xqa,DES:2020thz,Kovacs:2021edg}.

Taken together, these observations motivate the development of cosmological descriptions that explicitly account for departures from large-scale homogeneity and isotropy. Within General Relativity (GR), one can adopt two main approaches to \emph{physical cosmology} beyond the usual, statistically homogeneous and isotropic FLRW framework~\cite{Peebles:1980yev,Mukhanov:1990me}.
The first one is to describe inhomogeneities in terms of averaged geometric invariants, which produce kinematic backreaction terms in the averaged equations describing the cosmological dynamics~\cite{Buchert2000,Buchert2001,Buchert:2006ya,Wiltshire2007b,Green:2010qy,Green:2013yua,Clarkson:2011zq,Clarkson:2011uk,Sussman:2011na,BuchertRasanen2012,Racz:2016rss,Buchert:2018yhd,Buchert:2019mvq,Clifton:2019cep,Fanizza:2019pfp,Brunswic:2020bjx,Breton:2020puw,Lapi:2023plb,Giani:2024nnv,Gomes:2024iyq,Galoppo:2025rkd,Galoppo:2026rin,Raffai:2025ldr,Kolb:2005da}. The main limitation of this approach stems from the nonlinear nature of GR: in general, averaging the evolution equations governing inhomogeneities does not yield an effective homogeneous metric that is itself a GR solution. This failure of averaging to commute with the dynamics lies at the heart of the so-called fitting problem~\cite{Ellis:1984bqf,Ellis:1987zz,Clarkson:2011zq,Wiltshire:2011vy,Fleury:2025ykl}. An alternative way is to look for exact solutions to Einstein's equations which describe inhomogeneous cosmological models. Notable examples include generalizations of the FLRW framework, derived by introducing exact perturbations, which recover FLRW as a limit case~\cite{Clarkson:2010uz,Schander:2021pgt}, the Lema\^itre-Tolman-Bondi (LTB)~\cite{Lemaitre:1933gd,Tolman:1934za,Bondi1947}, the Szekeres~\cite{Szekeres1975,Szekeres:1975dx,Celerier:2024dvs} and the Swiss Cheese models~\cite{Einstein:1945id,Einstein:1946zz,Marra:2007pm,Biswas:2007gi}, which have been extensively studied in past literature~\cite{GarciaBellidoHaugboelle2008,Bolejko:2010eb,
Bolejko:2011jc,Bull:2011wi,MarraNotari2011,Zumalacarregui:2012pq,Hellaby:2012zz,Fleury:2013sna,Koksbang:2017arw,Koksbang:2019cen,Koksbang:2020zej,Koksbang:2020qry,Najera:2020jdm,Giani:2021gbs,Marra:2022ixf,Galoppo:2025hzy}.
A major limitation of existing exact inhomogeneous cosmological models is their restricted generality: they typically rely on perfectly isotropic matter sources, most commonly pressureless dust. Attempts to explain the late-time cosmological dynamics with backreaction effects or voids in inhomogeneous cosmologies have met only partial success until now, with a general difficulty in reproducing the observed magnitude of the DE density~\cite{Bolejko:2015gmk,Adamek:2014gva,Adamek:2017mzb,Macpherson:2018btl,Macpherson:2021gbh,Macpherson:2022eve,Camarena:2022iae,Oestreicher:2024mtw,Adamek:2024hme,Camarena:2025upt}\MP{\cite{Baumann:2010tm}}.

A more general framework to construct exact solutions describing inhomogeneities at large scales is provided by GR models sourced by anisotropic fluids~\cite{Herrera:1997plx}. In this setting, the fluid is assumed to be free of rotation, shear and acceleration. The anisotropy arises solely from the difference between the radial and tangential components of the pressure,~\textit{i.e.},~$p_{\parallel}(r,t)\neq p_{\perp}(r,t)$ (in this work, $t$ denotes the conformal time). Inhomogeneities are encoded in the radial dependence of the density profile $\rho(r,t)$ and the pressure components. Anisotropic fluid sources have been employed in a broad range of contexts, from regular black holes~\cite{Cadoni:2022chn} and the cosmological embedding of compact objects~\cite{Cadoni:2023lum,Cadoni:2023lqe,Cadoni:2024jxy,Cadoni:2024rri,deLima:2025bqf} to potential alternatives to CDM~\cite{Cadoni:2021zsl,Cadoni:2020izk,Cadoni:2018dnd}. The appeal of this description lies in its ability to accommodate radial inhomogeneities through the density and pressure profiles, while incorporating pressure anisotropy through the distinction between radial and tangential stresses. This offers a simple yet flexible framework that preserves the spherical symmetry of the metric despite the presence of anisotropic pressures. Previous attempts to develop a fully comprehensive and self-consistent framework for Anisotropic Fluid Inhomogeneous Cosmology (AFIC) have met limited success~\cite{Cadoni:2020jxe,Cadoni:2021zsl}. A recent development~\cite{Cadoni:2026ejk}, however, has provided a crucial step forward by deriving the most general spherically symmetric solution to Einstein's field equations sourced by an anisotropic fluid. 

At early times, when pressures are fully isotropic, AFIC reduces to FLRW cosmology. The same happens at late times, on sufficiently large scales, when AFIC naturally approaches the FLRW cosmological limit, ensuring that its predictions can be tested using standard observational methods based on cosmic expansion probes.
The main innovation of AFIC is its departure from the conventional perturbative assumption of small metric fluctuations, thereby allowing a fully relativistic description of the non-linear evolution of cosmic structures down to intermediate and small scales.
This new framework opens new avenues to reassess the viability of the AFIC solution to the DE puzzle. Such a renewed perspective is particularly timely in light of the data releases of the Dark Energy Spectroscopic Instrument (DESI)~\cite{DESI2024VI,DESIDR2II}, which favor a late-time emerging DE component over the standard cosmological constant hypothesis~\cite{DESI:2024aqx,DESI:2025fii,Scherer:2025esj,Tsujikawa:2013fta,DeFelice:2010aj,Amendola:2006kh,Amendola:2006we,Carroll:2004de,Carroll:1998zi,Odintsov:2026fyc,Odintsov:2025jfq,DOnofrio:2025cuk}. AFIC offers a natural setting in which to explore the emergence of such an effective DE component. In our framework, the term giving rise to DE does not enter the exact metric solution. It naturally emerges at late times as a large-scale effect induced by the inhomogeneous distribution of baryonic and dark matter. This mechanism offers an elegant solution to both the old and the new cosmological constant problems,~\textit{i.e.},~the hierarchically small value of $\Lambda$ compared to quantum field theory predictions~\cite{Weinberg:1988cp,Carroll:2000fy,Martin:2012bt}, and the coincidence problem~\cite{Zlatev:1998tr,Velten:2014nra}. 

The AFIC framework, built upon Ref.~\cite{Cadoni:2026ejk}, is parametrized by a single free input function: the Misner--Sharp (MS) mass~\cite{Misner:1964je,Akbarieh:2025vau}, which encodes the radial distribution of inhomogeneities at different redshifts. In this paper we show that, within this framework, an effective DE-like term emerges naturally in the large-scale limit, driven solely by inhomogeneities and pressure anisotropies in the total matter energy-momentum tensor. We further present a simple, exact realization of the model, in which the order of magnitude of the DE-like term, interpreted as an effective cosmological constant, is determined by the ratio between the characteristic mass of the inhomogeneities and the total mass quadrupole generated by the anisotropic matter distribution.

The remainder of the paper is structured as follows.
In~\cref{sect:afic} we present our inhomogeneous cosmological model. In~\cref{sect:eccls} we demonstrate the emergence of an effective DE in the large-scale limit,~\textit{i.e.},~in the FLRW limit of the model. In \cref{sect:simple} we introduce a simple, analytically tractable solution to Einstein's field equations. We interpret the emergent DE as an effective cosmological constant and provide an order-of-magnitude estimate of its present value within the model. In~\cref{conclusions} we state our conclusions and outline future observational prospects. For a more detailed discussion on the observational prospects of the AFIC predictions, we refer to our companion $Letter$~\cite{Cadoni:2026cyh}, where a first determination of the MS mass of the universe on Gpc scales is provided.

Additional technical calculations are reported in Appendix~\ref{App:Estimation_Q}. 

\section{Anisotropic Fluid Inhomogeneous Cosmology}
\label{sect:afic}

We adopt the following metric ansatz to describe the inhomogeneous cosmology~\footnote{We work in natural units and set $c = 1$.}
\begin{equation}
\label{eq:LineElementMain}
    \dd s^2 = a^2(t) 
    \left[
        -\ee^{\alpha(t, r)} \, \dd t^2 
        + \ee^{\beta(t, r)} \, \dd r^2 
        + r^2 \, \dd\Omega^2
    \right] .
\end{equation}
Here, \( \alpha(t, r) \) and \( \beta(t, r) \) are functions of the radial coordinate $r$ and the conformal time $t$, while \( a(t) \) is the scale factor. We effectively describe matter in the universe with an anisotropic fluid, which is given by the energy-momentum tensor (see,~\textit{e.g.}, Refs.~\cite{Herrera:1997plx, Ellis:1998ct, Horvat:2010xf} and references therein):
\begin{equation}
\label{eq:MomentTensorMain}
    T_{\mu \nu} 
    = (\rho + p_{\perp}) \, u_{\mu} u_{\nu} 
    + p_{\perp} \, g_{\mu \nu} 
    - (p_{\perp} - p_{\parallel}) \, w_{\mu} w_{\nu} \, ,
\end{equation}
where \( u_{\mu} \) and \( w_{\mu} \) satisfy the normalization conditions \( g^{\mu \nu} u_{\mu} u_{\nu} = -1 \), \( g^{\mu \nu} w_{\mu} w_{\nu} = 1 \), and \( u^{\mu} w_{\mu} = 0 \).

In the following, we encode inhomogeneities in the double dependence on $t$ and $r$ of $\rho$, $p_\parallel$ and $p_\perp$. We also adopt a simplifying assumption and encode anisotropies exclusively in the energy-momentum tensor, where they arise from the difference between the tangential and radial pressures, $p_{\perp}\neq p_{\parallel}$. 
We assume spherical symmetry and purely radial density profiles, a framework particularly suited to describing cosmic voids. Accordingly, the metric in~\cref{eq:LineElementMain} is not intended as a comprehensive description of the inhomogeneous Universe. Rather, it provides a tractable setting to investigate the interplay between matter and geometry in regimes that lie beyond the FLRW ansatz.

The effective description in terms of an anisotropic fluid finds support in several frameworks, including emergent and corpuscular scenarios~\cite{Verlinde:2016toy,Cadoni:2018dnd,Tuveri:2019zor,Giusti:2019wdx}.
A key ingredient in these scenarios is the existence of a characteristic scale at or below which the effective description becomes applicable. Previous studies have identified the typical size of a galaxy as the relevant scale. In Verlinde's emergent gravity, this scale marks the onset of the dominance of the thermal, volume-law contribution to the entropy associated with DE~\cite{Verlinde:2016toy}. In the corpuscular gravity framework and its anisotropic-fluid description~\cite{Tuveri:2019zor}, it instead corresponds to the crossover between Newtonian and non-Newtonian regimes. Contrary to these approaches, however, we here remain entirely agnostic about the physical origin of the departure from Newtonian dynamics at galactic scales~\cite{Frenk:2012ph,Strigari:2012acq,Lelli:2016cui,Lelli:2016zqa,Salucci:2018hqu}. Our framework naturally accommodates departures attributed either to (C)DM, as in the standard cosmological model, or to infrared modifications of GR.

Einstein's field equations following from the metric ansatz~\eqref{eq:LineElementMain} sourced by~\cref{eq:MomentTensorMain} can be written as~\cite{Cadoni:2020jxe} 
\begin{subequations} \label{eq:FirstFieldEq}
\begin{align} 
& \frac{\dot{a}}{a} \alpha' + \frac{\dot{\beta}}{r} = 0\, ;\label{eq:FirstFieldEqa}\\
& 3 \frac{\dot{a}^2}{a^2} + \frac{e^{\alpha - \beta}}{r^2} \left(-1 + e^{\beta} + r\beta' \right) + \frac{\dot{a}}{a} \dot{\beta} = 8 \pi G e^{\alpha} a^2 \rho\, ;\label{eq:rhoMain} \\
& \frac{\dot{a}^2}{a^2} e^{\beta - \alpha} + \frac{1 - e^{\beta} + r\alpha'}{r^2} \nonumber \\
&+ e^{\beta - \alpha} \left(-2 \frac{\ddot{a}}{a} + \frac{\dot{a}}{a} \dot{\alpha} \right) = 8 \pi G e^{\beta} a^2 p_{\parallel}\, ; \label{eq:FirstFieldEqc}\\
\label{eq:rhoConservMain}
& \dot{\rho} + \frac{\dot{a}}{a} \left(3 \rho + 3 p_{\parallel} + r p'_{\parallel} \right) = 0\, ; \\
\label{eq:p}
& p'_{\parallel} + \frac{\alpha'}{2} \left(\rho + p_{\parallel} \right) + \frac{2}{r} \left(p_{\parallel} - p_{\perp} \right) = 0\, ,
\end{align}
\end{subequations}
with dot and prime referring to derivation with respect to $t$ and $r$, respectively.

As a first step, we decompose the energy density $\rho$ and the radial pressure $p_{\parallel}$ in terms of a purely homogeneous FLRW contribution  and a term describing the inhomogeneities (denoted by the subscript $I$)
\begin{equation}
\label{eq:ass}
\begin{aligned}
  \rho(r, t) &= \tilde\rho_{\text{\tiny FLRW}}(t)+\rho_{\rm I}(r, t)\, , \\[4pt]
  p_{\parallel}(r, t) &= \tilde p_{\text{\tiny FLRW}}(t) + p_{\parallel \text{I}}(r, t)\, ,
\end{aligned}
\end{equation}
where $\tilde \rho_{\text{\tiny FLRW}}(t)$ and $\tilde p_{\text{\tiny FLRW}}(t)$ satisfy the standard Friedmann equations 
\begin{equation}
\label{eq:5}
\begin{aligned}
      &\frac{3 \dot{a}^2}{a^2} 
    = 8 \pi G a^2 \tilde\rho_{\text{\tiny FLRW}}\, ; \\[6pt]
   &-2 \frac{\ddot{a}}{a} + \frac{\dot{a}^2}{a^2} 
     = 8 \pi G a^2 \tilde p_{\text{\tiny FLRW}}\, ; \\[6pt]
&\dot{\tilde\rho}_{\text{\tiny FLRW}} + 3 \frac{\dot{a}}{a}
   \left( \tilde\rho_{\text{\tiny FLRW}} + \tilde p_{\text{\tiny FLRW}} \right) = 0 \, .
\end{aligned}
\end{equation}
No assumption on the equation of state describing matter is required. Consequently, our results are valid even beyond the standard CDM paradigm,~\textit{i.e.},~in any alternative scenario where DM is not assumed to be collisionless, pressureless dust~\cite{Ferreira:2020fam,deMartino:2020gfi,Angulo:2021kes}.

In the following, we focus on models that explicitly recover the FLRW metric at large scales. Throughout this work, we characterize this regime with the limit $r \to \infty$, which should be understood as a formal asymptotic prescription. Physically, it describes the regime in which the characteristic scales of matter inhomogeneities become much smaller than the scales under consideration, so that an effective homogeneous and isotropic description is expected to apply. The simplest way to achieve this is to set $\rho_{\rm I} \to 0$ and $p_{\parallel \text{I}}\to 0$ in the same limit. Accordingly, $\alpha, \, \beta \to 0$ so that \cref{eq:LineElementMain} reduces to the standard spatially flat FLRW metric~\footnote{We limit our analysis to the case of zero spatial curvature, as a non-negligible departure from flatness would break the consistency between CMB and baryon acoustic oscillation (BAO) data~\cite{Planck:2018vyg,Efstathiou:2020wem,Vagnozzi:2020dfn,CosmoVerseNetwork:2025alb}.}. As expected, when the contribution of inhomogeneities in \cref{eq:ass} is switched off, \cref{eq:FirstFieldEq} reduces to the Friedmann equations~\eqref{eq:5}.
The $r\to\infty$ limit is not the only regime in which our model reproduces FLRW cosmology. Indeed the latter can be recovered at all spatial scales whenever $\rho_{\rm I}=p_{\parallel \text{I}}=0$,~\textit{i.e.},~when inhomogeneities are negligible with respect to the FLRW background. This regime should be valid in the early stages of the cosmic expansion, where structure formation can be fully described in terms of linear perturbation theory, before the onset of the nonlinear growth. This important aspect highlights the fact that the AFIC model naturally reduces to the standard FLRW cosmology at early times, describing departures from the latter only at later times after recombination, namely when structure formation enters the nonlinear stage and $\rho_{\rm I}$ departs significantly from zero. In this regime, the FLRW background does not decouple anymore from that of the perturbations. The full gravitational dynamics is described by  \cref{eq:IntegrteFields0}, allowing one to overcome the difficulties of other approaches based on averaging and backreaction.

The system of equations~\eqref{eq:FirstFieldEq} is not closed and admits a general solution parametrized by a single free function~\cite{Cadoni:2026ejk}. Within our framework for inhomogeneous anisotropic cosmology, this function must provide a covariant description of cosmic matter distribution. The most natural choice is the MS mass~\cite{Misner:1964je}, a purely geometric quantity that provides an unambiguous quasi-local measure of the mass enclosed within a given radius. In spherically-symmetric spacetimes, it can be written as
\begin{equation}
\label{eq:Ms_total}
    M(r,t)
    = \frac{R}{2G}\left(1 - \nabla_\mu R  \, \nabla^\mu R\right),
\end{equation}
where $R \equiv a r$ is the areal radius~\footnote{In the context of cosmological coupling of compact objects~\cite{Cadoni:2023lum,Cadoni:2023lqe,Cadoni:2024jxy,Cadoni:2024rri,Cadoni:2026ejk}, the most natural choice for the free function is instead the mass of the local compact object $\mms^l$, defined as $M$ minus the cosmological background contribution [see Eq.~(7) in Ref.~\cite{Cadoni:2026ejk}]. Hence the difference in formulation between the system~\eqref{eq:IntegrteFields0} and the system~(6) of Ref.~\cite{Cadoni:2026ejk}.}.

After some manipulation (see Ref.~\cite{Cadoni:2026ejk} for details) the field equations can be recast as follows 
\begin{subequations}
\label{eq:IntegrteFields0}
\begin{align}
\label{IntegrteFields.a}
&\ee^{-\beta} = 
1 - \frac{2  G M(r,t)}{R}
+ \frac{8\pi  G}{3} R^2 \rho_{\text{\tiny FLRW}}\ee^{-\alpha}\,; \\
\label{IntegrteFields.b}
&\frac{\dot{a}}{a} \alpha' + \frac{\dot{\beta}}{r} = 0\, ;  \\
\label{IntegrteFields.cMain}
&\rho =  \frac{1}{4\pi a^3 r^2} 
\partial_r \!M\, ; \\
\label{IntegrteFields.dMain}
&p_{\parallel} =  - \frac{1}{4\pi \dot{a} a^2 r^3} 
\partial_t M\, ; \\
\label{IntegrteFields.e}
&p_{\perp} = 
p_{\parallel} 
+ \frac{r}{2} \partial_r p_{\parallel} 
+ \frac{1}{4} r \alpha' \left(\rho + p_{\parallel}\right)\, .
\end{align}
\end{subequations}
In \cref{IntegrteFields.a} $\rho_{\text{\tiny FLRW}}$ represents the total effective contribution of matter to the FLRW dynamics, including the inhomogeneities \footnote{Note that in Ref.~\cite{Cadoni:2026ejk} the equation corresponding to \cref{IntegrteFields.a} is written in terms of $\tilde\rho_{\text{\tiny FLRW}}$ rather than of $\rho_{\text{\tiny FLRW}}$. We can switch from one to the other equation just by the redefinition $\alpha \to \alpha + \ln(\tilde\rho_{\text{\tiny FLRW}}/\rho_{\text{\tiny FLRW}})$ followed  by a time reparametrization.}

\begin{equation}
  \rho_{\text{\tiny FLRW}}(t)= \tilde\rho_{\text{\tiny FLRW}}(t) +\rho_{\rm I}(r\to \infty, t)\,.
\end{equation}
We first highlight the essential role of anisotropies by discussing the isotropic limit, $p_{\parallel} = p_\perp$. In this case, \cref{IntegrteFields.e} becomes a differential constraint on $M$, which can therefore no longer be freely specified. With $p_\parallel \neq p_\perp$, instead, the system~\eqref{eq:IntegrteFields0} admits a unique solution only once $M$ is specified and appropriate boundary conditions are imposed~\cite{Cadoni:2026ejk}. The latter should be consistent with recovering homogeneity and isotropy at large scales, namely in the $r \to \infty$ limit. One can easily check that this is realized if the asymptotic MS mass is given by the FLRW contribution only,~\textit{i.e.},
\begin{equation}
\label{eq:MsFRW}
    M(r,\, a)
    \sim \frac{4\pi}{3} R^3 \rho_{\text{\tiny FLRW}}(a)\, \quad \text{for} \quad r \to \infty\, .
\end{equation}
The system~\eqref{eq:IntegrteFields0}, then, yields $\beta = \alpha = 0$ and $\rho=\rho_{\text{\tiny FLRW}}$, $p_{\parallel}= p_{\perp}=  p_{\text{\tiny FLRW}}=\tilde p_{\text{\tiny FLRW}}+p_{\parallel \text{I}}(r\to\infty , t)$, consistently with a spatially-flat FLRW solution. 

In order to be predictive, thus, the AFIC model only requires two inputs from observations: $(i)$ the late-time evolution and spatial distribution of matter fluctuations encoded in the MS mass~$M(r,\, a)$; $(ii)$ as in standard cosmology, the FLRW contribution to the total energy density, $\tilde\rho_{\text{\tiny FLRW}}$, and the equation of state for matter in the FLRW limit, $w_{\text{\tiny FLRW}} \equiv \tilde p_{\text{\tiny FLRW}}/\tilde\rho_{\text{\tiny FLRW}}$. The tangential pressure $p_{\perp}$ is, instead, determined by \cref{IntegrteFields.e} and is therefore an output of the model. If measurable, it would represent a smoking-gun test for our theoretical framework. 

We conclude this section by comparing the AFIC framework with the LTB model, one of the most extensively studied and widely used frameworks to describe inhomogeneous cosmologies. The metric parametrizations are fundamentally different. Indeed, identifying the LTB areal radius with the AFIC radial coordinate, namely $R(r,t) = a(t)r$, would reduce the LTB evolution equation, together with the covariant energy-momentum conservation, to the FLRW equations, thereby enforcing homogeneity and isotropy. Thus, the LTB model cannot be recovered as a limit case of the AFIC framework. This is a consequence of the fact that the LTB model assumes an isotropic source. Conversely, AFIC is sourced by an anisotropic fluid and therefore encompasses a broader class of cosmological configurations. Moreover, a general LTB solution is specified by three arbitrary functions of the radial coordinate: the mass profile $M(r)$, the spatial curvature function $E(r)$, and the bang-time function $t_B(r)$, which allows different spacetime regions to begin expanding at different times. In contrast, our framework is governed by a single free function, the MS mass, depending on both $r$ and the conformal time $t$. This dependence provides the key input for connecting the inhomogeneous matter distribution to its cosmological evolution, thereby rendering the framework predictive. A final important distinction concerns the Copernican principle. In the LTB framework, isotropy is realized around a preferred central worldline, which is physically distinguished from all other observers. This feature violates the Copernican principle, even when the center, $R=0$, is regular and free of curvature singularities. The underlying reason can be traced to the Ehlers–Geren–Sachs (EGS) theorem~\cite{EGS1968}, according to which, for pressureless dust and, more generally, isotropic fluids, isotropy about every fundamental observer implies spatial homogeneity. This result does not generally extend to anisotropic fluids. Consequently, relaxing the assumption of isotropic pressure opens up the possibility of constructing genuinely inhomogeneous cosmologies that preserve the Copernican principle, without introducing a physically privileged observer or a distinguished central worldline. We will briefly return to this point in~\cref{sect:simple}.

\section{Emergence of a dark energy-like component at large scales}
\label{sect:eccls}

A remarkable feature of the AFIC framework is the emergence of an effective DE-like term at large scales, driven by the late-time inhomogeneous distribution of cosmic matter. This mechanism is \emph{independent} of the specific functional form of the MS mass: it only depends on its asymptotic behavior. Importantly, this component is not necessarily restricted to a cosmological constant, but can exhibit a  dynamical dependence on the scale factor.

The key to this mechanism lies in splitting the MS mass into two distinct contributions, as a consequence of the decomposition introduced in \cref{eq:ass}: 
\begin{equation}
M(r,\, a) = M_\text{\tiny FLRW}\left(R,\tilde \rho_{\text{\tiny{FLRW}}}\right) + M_\text{I}(r,\, a)\, ,
\end{equation}
where $M_\text{\tiny FLRW}=\left(4\pi/3\right) R^3 \tilde \rho_{\text{\tiny{FLRW}}}$ accounts for the FLRW background contribution. Since we are mostly interested in late-time cosmology, we here assume this background to be dominated by pressureless matter with density $\rho_\mathrm{M}(a)$. 
$M_{\text{I}}(r,\, a)$, instead, encodes the contribution arising from inhomogeneities in the matter distribution. If the latter does not vanish in the large-scale FLRW limit ($r\to \infty$), we must have in this regime
\begin{equation}
\label{eq:mlimit}
    M(r, \, a)
    \sim \frac{4\pi}{3} R^3 {\rho}_{\text{\tiny FLRW}}=\frac{4\pi}{3} R^3 \left[\tilde{\rho}_{\text{\tiny FLRW}}+\lambda \, g(a)\right]\,  ,
\end{equation}
where $\lambda$ is a constant with dimensions of $\text{length}^{-4}$ and $g$ is a function of $a$ only, to ensure compatibility with the requirement of large-scale homogeneity and isotropy. Although the limit $r\to\infty$ of $M$ is formally divergent, a physically meaningful characterization of the cosmological mass at present time ($a=1$) is provided by its value at the apparent horizon, $M^{\rm AH}$~\cite{Hellaby:2006cj}. For a spatially flat FLRW universe, the apparent-horizon radius is $R_{\rm AH}=H^{-1}$. Considering that at the apparent horizon the second term in \cref{eq:Ms_total} is zero, we get $M^{\rm AH}=1/(2GH)$ . This quantity provides a natural connection between the AFIC predictions and cosmological observations~\cite{Cadoni:2026cyh}. 
In this setting, the contribution to the MS mass coming from inhomogeneities, encoded in $\lambda$, can be interpreted as an effective DE component with a time-dependent density and, correspondingly, an effective dynamical equation of state $w_{\rm DE}=w_{\rm DE}(a)$. In particular, the present-day value $w_{\rm DE}(a=1)$ need not coincide with $-1$, nor does our framework require the effective DE component to behave as a cosmological constant at all epochs. This possibility is particularly interesting in light of recent DESI results, which have reported a preference for a time-evolving DE equation of state~\cite{DESI2024VI,DESIDR2II,DESI:2024aqx,DESI:2025fii,Scherer:2025esj}. 

While in our companion $Letter$~\cite{Cadoni:2026cyh} we explore this perspective by providing the first observational constraints on the MS mass on Gpc scales, in what follows we focus on the simplest case $g(a) = 1$. This choice allows us to draw conceptual comparisons between the standard cosmological constant $\Lambda$ and the effective cosmological-constant-like term emerging in AFIC. The conceptual difference between the two becomes evident when considering the impact on the spacetime geometry. A static energy distribution related to the standard cosmological constant $\Lambda$ enters the MS mass with a term $R^3 \Lambda/(6G)$, generating a contribution $\Lambda R^2/3$ in the metric function, yielding the de Sitter spacetime. Conversely, AFIC provides an entirely new mechanism. Treating \cref{eq:mlimit} as the leading term in the asymptotic expansion and imposing $\alpha,\, \beta \to 0$ in this limit, as required to recover FLRW cosmology, we find that the contribution proportional to $\lambda$ cancels out~\footnote{Despite this, $\lambda$ generically enters the subleading terms in the asymptotic expansion of the metric functions. Its value is therefore determined by the functional form of $M_{\text{I}}(r,\,a)$ and the parameters entering it.}. Thus, $\lambda$ does not enter directly the local geometry of the inhomogeneous universe and it is entirely ascribable to the inhomogeneous matter distribution. Additionally, a constant $\lambda$ is expected to become dominant at late times, due to the dilution of the matter density as $\rho_{\rm M}\propto a^{-3}$. Taking the limit $r\to\infty$ at different redshifts provides a potential way to identify the redshift $z_{\rm eq}$ at which accelerated expansion should set in. 

Despite their different cosmological evolution, however, $\lambda$ and $\rho_{\rm M}$ both contribute to the MS mass through terms scaling as $R^3$. This cubic scaling is a generic prediction of FLRW cosmology. Directly testing it on scales of $ \gtrsim 200$ Mpc is challenging, primarily because gravitational lensing probes do not extend to such large radii~\cite{Wambsganss:1998gg,Refregier:2003ct,Munshi:2006fn,Fu:2007qq,Hoekstra:2008db,Bartelmann:2010fz,Rasanen:2014mca,Liao:2017yeq,Qi:2018atg,Qi:2018aio,Wagner:2018jxp,Li:2019rns,Liao:2019hfl,Cao:2019kgn,Kumar:2020ole}. Nevertheless, it has been indirectly tested through CMB and LSS observations~\cite{Camarena:2021mjr,Aluri:2022hzs}.

We conclude this section with some remarks clarifying the physical interpretation and scope of our results. First, our approach does not identify the mechanism through which inhomogeneities generate an effective DE component. Rather, the AFIC framework establishes that an inhomogeneous matter distribution can give rise to such an emergent term in the large-scale FLRW regime. 
This interpretation is consistent with our description in terms of an anisotropic fluid, which becomes isotropic in the homogeneous limit. The fluid should therefore be regarded as an effective, simplified description of the actual late-time inhomogeneous gravitational clustering properties on small and intermediate scales. 
On the other hand,  although it appears rather unnatural, we cannot exclude the conventional interpretation of the term $\lambda g(a)$: namely, that it describes an exotic dark energy component already present in the early Universe, but which becomes manifest only at late times. Within AFIC, it emerges as an extensive contribution to the asymptotic limit of the MS mass.
Second, the mechanism of the emergence of a cosmological-constant-like term differs fundamentally from conventional approaches based on an FLRW background, namely spatial averaging of inhomogeneities or backreaction. In the AFIC framework, instead, we start from an exact solution to Einstein's equations that consistently incorporates the cosmological background, local inhomogeneities, and their gravitational interplay. The cosmological-constant-like term emerges only in the $r\to\infty$ limit, without being imposed as a background ingredient.

\section{A simple model for the emergence of a cosmological constant from inhomogeneities}
\label{sect:simple}

Ref.~\cite{Cadoni:2026ejk} derived the most general unique solution to the system~\eqref{eq:IntegrteFields0}, which is not amenable to a simple closed-form representation. Here, we focus on a particularly simple realization of AFIC for which an explicit closed-form solution can be obtained. We adopt the following redefinition of the metric function $\alpha$,  
\begin{equation}
\label{eq:red}
\alpha \equiv \gamma+ \ln\frac{\tilde\rho_{\text{\tiny FLRW}}}{\lambda}, 
\end{equation}
where $\lambda$ is a constant, and we assume $M$ to be a function of $R$ only
 \begin{equation}
\label{eq:ms1}
   M=M(R) \, .
\end{equation}
At first sight, this assumption may appear restrictive even in standard FLRW cosmology, where $M$ is given by \cref{eq:MsFRW}. However, it is justified within the quasi-static approximation, which can be applied at any cosmological epoch. Indeed, for a region much smaller than the Hubble scale and over a time interval $\delta t\ll H^{-1}$, the cosmological density can be regarded as approximately constant, $ \rho_{\text{\tiny FLRW}}(t)\approx \rho_{\text{\tiny FLRW}}(t_0)=\text{constant}$, so that $M\propto R^3\rho_{\text{\tiny FLRW}}(t_0)$. We extend this approximation to our inhomogeneous cosmology, assuming that the characteristic scales of the inhomogeneities are sufficiently smaller than the Hubble radius for their evolution to be effectively decoupled from the cosmic expansion. Thus, \cref{eq:ms1} is understood to hold at least within this quasi-static regime.

Since $M$ depends only on $R$, consistency requires $\alpha$ and $\gamma$ to share the same dependence. Consequently, the system~\eqref{eq:IntegrteFields0} reduces to 
\begin{subequations}
\label{eq:IntegrteFields01}
\begin{align}
&\ee^{-\beta(R)} = 
1 - \frac{2  G M(R)}{R}
+ \frac{8\pi G }{3} R^2   \lambda \, \ee^{-\gamma(R)}\,; \\
%
&\frac{\dd}{\dd R}\left(\beta+\gamma\right)=0\, ;  \\
\label{IntegrteFields.cMainSimple}
&\rho = -p_{\parallel}= \frac{1}{4\pi R^2} \frac{\dd M}{\dd R}\, ; \\
&p_{\perp} = 
p_{\parallel} 
+ \frac{R}{2} \frac{\dd p_{\parallel} }{\dd R}\, . \label{IntegrteFields.dMainSimple} 
\end{align}
\end{subequations}
Through a redefinition of the conformal time, the solution of the above system takes the form
\begin{equation}
\label{eq:simplesol}
\begin{split}
    &\ee^{-\beta(R)} = \ee^{\alpha(R)} 
    =\frac{1}{2}\left[1 - \frac{2  G M(R)}{R}\right.\\
&\left. + \sqrt{\left(1 - \frac{2  G M(R)}{R}\right)^2+\frac{32\pi G}{3}\lambda R^2}\right]\,.
\end{split}
\end{equation}
The explicit form of the solution~\eqref{eq:simplesol} makes it possible to investigate the regularity of the resulting geometry directly in terms of the MS mass profile. In particular, a potential source of curvature singularities is the center $R=0$, where the metric functions and the matter variables must remain sufficiently regular. It can be shown that curvature invariants remain finite if $M= C R^3+ \mathcal{O}({R}^4)$ near $R = 0$ \footnote{Except possibly at $a=0$, where the usual cosmological singularity may occur.}.
Through \cref{IntegrteFields.cMain}, this implies that both the density and the pressures approach finite constants as $R\to 0$. Exploiting the freedom in $M$, we adopt this behavior throughout the following analysis. 

The regularity of the origin also has an interesting geometrical implication: since $R=0$ does not correspond to a singular or otherwise distinguished physical point, the coordinate origin $r = 0$ can be assigned to any location in the universe. However, as discussed in \cref{sect:afic}, this property does not automatically guarantee that the Copernican principle holds in our inhomogeneous cosmological model. The issue stems from the isotropy assumption built into the metric ansatz~\eqref{eq:LineElementMain}, together with the anisotropic nature of the fluid. In particular, it remains unclear whether the EGS theorem extends to cosmologies sourced by anisotropic fluids. If such an extension holds, the Copernican principle would likewise be incompatible with \cref{eq:LineElementMain} in the presence of an anisotropic fluid.

\subsection{Emergence of a cosmological constant}

We now consider the expansion of \cref{eq:simplesol} at large $R$, which describes the transition from an inhomogeneous universe to the FLRW asymptotic regime. We begin by expanding the MS mass. From \cref{eq:mlimit}, the mechanism yielding an emergent cosmological constant is related to a leading contribution proportional to $R^3$ in the MS mass. This term ensures that the geometry correctly reduces to the FLRW one for $R \to \infty$. Additionally, one can show that quadratic and linear terms in $R$ must vanish to maintain consistency with the correct asymptotic behavior. Therefore, we must require, in the limit $R \to \infty$,  
\begin{equation}
\label{eq:expmass}
    M(R) =\sigma R^3 + M_0+  \frac{M_{-1}}{R}+  {\cal{O}}(R^{-2}) \, ,
\end{equation}
where $\sigma$,  $M_0$ and $M_{-1}$ are some constants.
Rather than in a void (as many inhomogeneous models do), just for this example model, we assume that the origin of our metric is at the centre of an overdensity, which we identify  with a galaxy with mass $M_0$ (see discussion below). 
Note that $\sigma$ contains both the matter contribution and that due to $\lambda$. The former is not negligible (about $0.3$ of the total budget), but since we are just estimating the order of magnitude of the cosmological constant we assume
$\sigma \approx 4 \pi \lambda/3$.
Substituting this expansion into \cref{eq:simplesol}, one obtains, up to order $\mathcal{O}(R^{-4})$,
\begin{equation}
\label{multipole}
    \ee^{-\beta(R)} = 1 - \frac{k_{-3}}{R^3} -\frac{k_{-4}}{R^4} +\mathcal{O}(R^{-5})\, ,
\end{equation}
with
\begin{equation}
\label{expansion}
\begin{aligned}
    k_{-3} &\equiv \frac{M_0}{\sigma}\, ; \qquad k_{-4} \equiv \frac{M_{-1}}{\sigma}\, . 
\end{aligned}
\end{equation}
As mentioned in the previous section, the $\sigma R^3$ contribution in the MS mass does not contribute directly to the metric functions. Comparing this leading term in $M$ with the one expected from a standard cosmological constant,~\textit{i.e.},~$M_\Lambda = R^3 \Lambda/(6G)$ and using \cref{expansion}, we find the expression of the effective emergent cosmological constant \footnote{We note that $\sigma$ contains not only the cosmological constant term, but also the matter contribution $\rho_{\rm M}$ [see \cref{eq:mlimit}], which however is expected to become subdominant [see the discussion below \cref{eq:mlimit}].}
\begin{equation}{\label{eq:ecc1}}
    \Lambda_{\rm eff} = 6G \frac{M_0}{k_{-3}} = 6G \frac{M_{-1}}{k_{-4}}\, .
\end{equation}
%

%
%

We also note that \cref{multipole} has the structure of a multipole expansion with vanishing monopole and dipole contributions. At the present cosmological epoch, $a=1$, this expansion can be interpreted in terms of the Newtonian potential $\Phi= (\ee^{-\beta}-1)/2$ associated with the matter inhomogeneities. While the vanishing of the dipole contribution indicates that we have correctly chosen the origin of our radial coordinate system in the center of mass, the vanishing monopole contribution has a deeper physical significance. Although $M_0$ appears explicitly in the MS mass expansion~\eqref{eq:expmass} (hence it could be measured by local observations), it does not generate a monopole contribution to the asymptotic gravitational potential. Physically, this reflects the absence of a localized source, with matter inhomogeneities distributed throughout spacetime. The higher-order multipole coefficients, beginning with the quadrupole-order term, are generally nonvanishing and depend on both $\sigma$ and the coefficients $M_i$ in the expansion of $M$. This opens up the possibility of constraining the value of the effective cosmological constant through the asymptotic multipole structure of the inhomogeneities, provided the relevant coefficients can be independently determined. Let us thus compare our solution \eqref{eq:expmass} with the usual multi-polar expansion of the $g_{tt}(\ell)$ component of the metric, with $\ell$ a reference scale. Since the monopole and dipole terms, respectively proportional to $\ell^{-1}$ and $\ell^{-2}$, are zero in our case, we have
 \begin{equation}
 \label{eq:expmass2}
 \begin{split}
    \ee^{-\beta(\ell)} = & 1 -G\frac{Q_{ij}n^in^j}{\ell^{3}}-\frac{G}{3}\frac{Q_{ijk}n^in^jn^k}{\ell^{4}}+ {\cal{O}}(\ell^{-5})\, ,
\end{split}
\end{equation}
where $Q_{ij}$, $Q_{ijk}$ are the quadrupole and octupole contributions, respectively, while $n^i$ is a unit radial vector. Comparing \cref{multipole,eq:expmass2} gives the expression of the effective cosmological constant in terms of $M_0$, $M_{-1}$ and of the quadrupole and octupole momenta generated by the inhomogeneities 
\begin{equation}{\label{eq:ecc}}
    \Lambda_{\rm eff} = \frac{6 M_0}{Q_{ij} n^i n^j} = \frac{18 M_{-1}}{Q_{ijk} n^i n^j n^k}\, .
\end{equation}
%

%
The expansion \eqref{eq:expmass2} is obtained in the quasi-static approximation at the present epoch ($a=1$). At first sight, extracting a cosmological constant in a regime where cosmic expansion is effectively frozen may seem paradoxical. Yet, this is precisely the distinctive feature of our mechanism. Local matter inhomogeneities do not determine $\Lambda_{\rm eff}$ directly through the metric potential, whose multipole expansion \eqref{multipole} contains no $\ell^2$ term, nor do they directly fix the constant mass contribution $M_0$. Rather, they determine the multipole coefficients of the gravitational potential. Since both $\lambda$ and $M_0$ enter $M$, these coefficients encode global cosmological information, including the value of $\Lambda_{\rm eff}$, in local gravitational observables.

\subsection{Estimation of the effective cosmological constant from observable quantities}

We now estimate the order of magnitude of the emergent cosmological constant using \cref{eq:ecc}. Several limitations of our model preclude a more precise determination. First, the model does not provide a fully realistic description of matter inhomogeneities, as it neglects their anisotropic structure, which is parametrized by nonvanishing higher-order multipole coefficients $k_i$. Second, our analysis relies on the asymptotic expansion \eqref{eq:expmass}, which is valid only at sufficiently large $R$. Third, extracting the mass quadrupole $Q_{ij}$ and the constant mass contribution $M_0$ from observations requires detailed knowledge of the matter distribution, including its anisotropies. Accordingly, our estimate of the emergent cosmological constant should be regarded as semi-quantitative, providing an order-of-magnitude assessment rather than a precise determination.

Before using \cref{eq:ecc}, we need to estimate $M_0$ (the subleading coefficient $M_{-1}$) and the mass quadrupole $Q_{ij}$ (resp., the octupole $Q_{ijk}$). These quantities pose distinct observational challenges. While the multipole momenta enter explicitly \cref{multipole} and therefore admit an operational definition in terms of the gravitational potential, $M_0$ and $M_{-1}$ do not appear in the metric function, and hence they cannot be determined by direct cosmological observations. In particular, \cref{IntegrteFields.cMainSimple,IntegrteFields.dMainSimple} show that $M_0$ does not contribute to the cosmological density or pressures. Thus, although it characterizes local inhomogeneities and enters the determination of $\Lambda_{\rm eff}$ through \cref{eq:ecc}, it does not directly affect the large-scale cosmological dynamics. Its value must therefore be inferred from local measurements of the gravitational field generated by matter inhomogeneities. 

The absence of a preferred scale in the matter distribution, however, prevents a unique determination of $M_0$ from the large-scale structure alone: cosmic structures form over a broad range of scales, determined by the primordial power spectrum and subsequent gravitational evolution. The determination of $M_0$ must therefore be tied directly to the range of validity of our effective description of matter and its inhomogeneities as an anisotropic fluid. The natural scale for fixing $M_0$ is the transition scale at which this effective description becomes applicable, which we identify with the size of a typical galaxy. Within the standard $\Lambda$CDM model, $M_0$ includes both baryonic and dark matter. The same value of $M_0$ can be adopted in scenarios involving IR modifications of gravity~\cite{Milgrom:1983ca,Milgrom:1983pn,Clifton:2011jh,Capozziello:2011et,Verlinde:2016toy,Cadoni:2017evg,Cadoni:2018dnd,Tuveri:2019zor,Cadoni:2020izk}.

We next estimate the mass quadrupole $Q_{ij}$ associated with matter inhomogeneities. Spherical symmetry implies $Q_{ij}=0$, whereas a nonvanishing quadrupole requires an anisotropic mass distribution. 
Its determination is observationally challenging and further complicated by the conceptual issues concerning the Copernican principle discussed at the end of \cref{sect:afic}. 

We can nonetheless provide an estimate of the order of magnitude of $Q_{ij} n^i n^j$ considering the statistical distribution of cosmic structures, encoded in the linear matter power spectrum. First, we can exploit the freedom in arbitrarily choosing the origin of the radial coordinate $r$ [see the discussion below \cref{eq:simplesol}], and fix it to be the center of mass of an overdensity with mass $M_0$. We then consider the surrounding distribution of (baryonic and dark) matter with total mass $M$ with linear size $\mathcal{R}$ and background density $\tilde\rho_{\text{\tiny FLRW}} \sim \rho_m \sim 4 \cdot 10^{10} M_\odot \text{Mpc}^{-3}$~\cite{Planck:2018vyg}{\footnote{The  mass $M$ must not be confused with the mass $M_0$. The former is the total mass of our universe enclosed in a sphere of radius $\mathcal{R}$, whereas the latter is the mass of the overdensity, which in \cref{eq:expmass} plays the role of a mass monopole.}. Since $\mathcal{R}$ is not determined a priori within our framework, we adopt the interval $10^2 \, \text{Mpc} \lesssim \mathcal{R} \lesssim 10^3 \, \text{Mpc}$, which spans from the transition regime toward statistical homogeneity and isotropy to scales at which this regime is already well established. The details of the estimate of the order of magnitude of $Q_{ij} n^i n^j$ are presented in Appendix~\ref{App:Estimation_Q} and give 
\begin{equation}
    1.4 \cdot 10^{173} 
    \; m_{\rm P}\,\ell_{\rm P}^2 \, \lesssim\; Q_{ij} n^i n^j 
    \;\lesssim\; 4.5 \cdot 10^{176} \, m_{\rm P}\,\ell_{\rm P}^2\,,
    \label{Q}
\end{equation}
corresponding to the extrema of the considered interval of $\mathcal{R}$. As for $M_0$, we identify it as the total mass (baryonic $+$ dark matter) of a typical galaxy
\begin{equation}
    M_0 \approx [10^{11},\, 10^{12}] \, M_{\odot} \sim 9 \cdot [10^{48}, \, 10^{49}] \, m_{\rm P}\, . 
    \label{M0}
\end{equation}
This choice provides the largest-scale direct mass determination currently available that does not rely on a specific cosmological model (see also Ref.~\cite{Cadoni:2026cyh} for a discussion of model-independent
mass determination on large scales). 

Using \cref{Q,M0} into \cref{eq:ecc}, we get 
\begin{equation}
    1.2\cdot 10^{-127}\;\ell_{\rm P}^{-2}
    \;\lesssim\;
    \Lambda_{\rm eff}
    \;\lesssim\;
    3.9\cdot 10^{-123}\;\ell_{\rm P}^{-2}\,,
    \label{eq:Lambda_eff_range}
\end{equation}
where the lower (upper) bound corresponds to 
$M_0 = 10^{11}\,M_\odot$, $\mathcal{R}=10^3\,\text{Mpc}$ 
($M_0=10^{12}\,M_\odot$, $\mathcal{R}=10^2\,\text{Mpc}$). 
The upper end of this range falls within one order of magnitude of the 
observed value $\Lambda_{\rm obs}\approx 2.9\cdot 10^{-122}\,\ell_{\rm P}^{-2}$~\cite{Planck:2018vyg}, 
corresponding to a galaxy of mass $M_0\sim 10^{12}\,M_\odot$ 
averaged over a scale $\mathcal{R}\sim 10^2\,\text{Mpc}$.

While these estimates are obtained within a simplified model and 
should not be regarded as an exact prediction, they reduce the 
discrepancy with the observed value of the cosmological constant to at most a few orders of 
magnitude. 
Our result strongly indicates that -- within standard GR -- the origin of cosmic acceleration can be traced back to 
the actual, late-time inhomogeneous distribution of matter. For a detailed discussion on the actual testability of such an intriguing hypothesis, we refer to our companion $Letter$~\cite{Cadoni:2026cyh}.

\section{Conclusions}
\label{conclusions}

In this paper, we have introduced a new framework for inhomogeneous cosmology based on GR sourced by an anisotropic fluid. The resulting system admits exact solutions to Einstein's equations, recovering the standard FLRW cosmology both at large distances and for early-time cosmology, while consistently incorporating the nonlinear dynamics of matter inhomogeneities. Remarkably, the entire construction is governed by a single free input function, the MS mass profile $M(r,a)$, which encodes the spatial distribution of matter and its evolution throughout cosmic history. 
The AFIC framework is particularly well suited for investigating the transition from a matter-dominated  universe to accelerated expansion. Indeed, a central result of our work is the identification of a mechanism through which a DE-like component emerges  in the FLRW regime from the inhomogeneous distribution of baryonic and dark matter.
The link between cosmic acceleration and nonlinear structure formation we have found in our AFIC model is quite intriguing also in view of the historic development of modern cosmology. Cosmological $N$-body simulations for LSS formation were developed within the  CDM framework in the eighties of last century \cite{Blumenthal:1984bp,Efstathiou:1985zz,Davis:1985rj,White:1987zz,Frenk:1988zz,Maddox:1990hb}. It turned out that a $\Lambda$-term was needed to explain the LSS of the universe~\cite{Efstathiou:1990xe,Ostriker:1995su,KraussTurner1995}, anticipating by several years the detection of cosmic acceleration in type Ia Supernovae data~\cite{Perlmutter:1997zf,Riess:1998cb,Perlmutter:1998np}.

For simplicity, we focused on the realization of DE in terms of an emergent cosmological-constant-like term. In our model, the latter is generated if, in the limit $r\to\infty$, the MS mass profile $M(r,a)$ approaches the extensive proper-volume scaling given in \cref{eq:mlimit}. Remarkably, while this extensive contribution emerges in the FLRW limit, it does not explicitly appear in the exact metric describing the inhomogeneous spacetime. This provides a novel mechanism for the emergence of an effective cosmological constant, without resorting to the averaging or backreaction procedures commonly invoked in attempts to explain the cosmological effects of inhomogeneities. 
To illustrate this mechanism, we have also constructed a simple, exactly solvable realization of the AFIC model. In the FLRW limit, the effective cosmological constant is determined by the ratio between the constant mass mode $M_0$, associated with matter inhomogeneities, and the total mass quadrupole $Q_{ij}$ generated by their anisotropic distribution. Using this relation, we have obtained an interval of estimates that fall $5$ to $1$ orders of magnitude within the observed value, representing a substantial improvement over the present status of the cosmological constant problem. 

From a phenomenological perspective, however, two important questions remain open: the range of validity of this simplified model and the accurate determination of $M_0$ and $Q_{ij}$ from astrophysical observations. Addressing these issues is essential to assess the robustness of the mechanism and its ability to reproduce the observed value of the cosmological constant.

Our results raise several conceptual and phenomenological questions that deserve further investigation. At the conceptual level, an important issue concerns the use of an anisotropic fluid to describe cosmological inhomogeneities. This choice is motivated by the generality of the framework and by the widespread use of anisotropic fluids in the description of astrophysical compact objects. Nevertheless, it remains to be understood to what extent such a fluid provides an adequate effective description of the inhomogeneous matter distribution in cosmology. Closely related is the question of the Copernican principle in inhomogeneous cosmologies sourced by anisotropic fluids, and of the conditions under which the EGS theorem applies in this setting. Another important direction concerns the extension of our effective emergent DE mechanism beyond a strictly constant cosmological term. In particular, an appropriate choice of the mass profile $M(r,a)$ may lead to a time-dependent effective DE density and an equation of state different from $p_{\rm DE}=-\rho_{\rm DE}$. This possibility is particularly interesting in light of recent DESI results~\cite{DESI2024VI,DESIDR2II}, which have reported a preference for evolving DE over a cosmological constant for certain combinations of cosmological datasets. Understanding how our framework can accommodate such an evolution, while simultaneously addressing the cosmological coincidence problem, represents an important direction for future work.

From a phenomenological perspective, a primary challenge is to determine the MS mass profile $M(r,a)$ directly from observations, with no assumptions on the underlying cosmology. This requires detailed information on the formation and evolution of cosmic structures, potentially encoded in the cross-correlation signal between spectroscopic and photometric galaxy surveys, the gravitational lensing of the CMB, and the late ISW effect. Such a reconstruction would not only provide the input needed to make the AFIC framework fully predictive, but would also allow us to determine its asymptotic behavior as $r\to\infty$ at different epochs. As shown in our companion $Letter$~\cite{Cadoni:2026cyh}, the MS mass in the $r\to\infty$ limit  can be measured, allowing us to predict the correct present-day effective DE density (or, in the more general case, its equation of state). If high-precision multi-probe observations show that $M(r, z)$ matches the observed expansion history, late-time cosmic acceleration
will be proved to be a structural effect of standard GR.
If not, the AFIC framework will help point to where and how alternative extended theories must step in.

\vspace{3mm}

\section{Acknowledgments}
We are very grateful to Julian Adamek, Rodrigo Calderón, Matteo Cinus, Marco Galoppo, and especially Nils Sch\"oneberg for their numerous suggestions that definitely improved our work. We also thank Eric V. Linder for many insightful discussions. We acknowledge support from the \href{https://www.sgws-community.eu}{Sardinian GW Science (SGWS) Community}: a joint program between the University of Cagliari and the Gran Sasso Science Institute (GSSI) promoted and coordinated by the cultural association \href{https://www.associazioneideas.eu}{IDeAS}. APS acknowledges financial support by the MUR FIS2 Advanced Grant ET-NOW (CUP:~B53C25001080001). MC, RM, MP, MPit, APS, and IU acknowledge support by the INFN TEONGRAV specific initiative.

\onecolumngrid
\begin{appendix}
    \section{Statistical estimation of the quadrupole moment on large scales}
    \label{App:Estimation_Q}

In this appendix we provide a statistical estimate of $Q_{ij}n^i n^j$ in \cref{eq:ecc} 
on scales $\mathcal{R}$ by relating it to the matter power spectrum. As explained in the main text, we select values of $\mathcal{R}$ in an interval covering both the transition regime toward statistical homogeneity and isotropy and scales at which the Universe is already well within that regime, while keeping the linear matter power spectrum a valid description throughout. As we shall see at the end of the Appendix, a statistical estimate of the quadrupole moment reduces to evaluating $\langle Q_{ij}Q^{ij}\rangle$.  

Denoting by $\rho$ the \emph{total} density characterizing the interior of the volume with linear size $\mathcal{R}$ and writing it in terms of the mean density $\tilde \rho_{\text{\tiny FLRW}}$, sourcing the FLRW background, and the density contrast $\delta$ as $ \rho = \tilde \rho_{\text{\tiny FLRW}} (1 + \delta) $, we can define the quadrupole moment tensor as (bold letters will denote three-dimensional vectors)
\begin{equation}
\begin{split}
    Q_{ij} &= \int_{|\textbf{x}| < \mathcal{R}} \dd^3 x \, \rho(\textbf{x}) \, \left[3 x_i x_j - |\textbf{x}|^2 \delta_{ij} \right] = \tilde \rho_{\text{\tiny FLRW}} \int_{|\textbf{x}| < \mathcal{R}} \dd^3 x \, \delta(\textbf{x}) \left[3 x_i x_j - |\textbf{x}|^2 \delta_{ij} \right]\, ,
\end{split}
\label{eq:QuadrupoleMomentum_General}
\end{equation}
where in the second step, we used the fact that $\tilde \rho_{\text{\tiny FLRW}}$ pertains to a spherically-symmetric distribution of matter. 

We then express the density contrast in momentum space through its Fourier transform
\begin{equation}
    \delta(\textbf{x}) = \int \frac{\dd^3 k}{\left(2\pi \right)^3} \, \delta(\textbf{k}) \, \ee^{\ii \textbf{k} \cdot \textbf{x}}\, ,
    \label{eq:delta_FourierTransform}
\end{equation}
and define the matter power spectrum as
\begin{equation}
    \langle \delta(\textbf{k}) \delta(\textbf{k}')\rangle = \left(2\pi \right)^3 \delta^3 \left(\textbf{k} - \textbf{k}' \right) P(k)\, ,
    \label{eq:PowerSpectrum_Definition}
\end{equation}
where we assumed $P(\textbf{k}) = P(k)$ for a statistically homogeneous and isotropic field, which on scales $\mathcal{R}$ is a reasonable assumption.

Plugging \cref{eq:delta_FourierTransform} into \cref{eq:QuadrupoleMomentum_General} gives 
\begin{equation}
    Q_{ij} = \tilde \rho_{\text{\tiny FLRW}} \int \frac{\dd^3 k}{(2\pi)^3} \delta(\textbf{k}) \mathcal{W}_{ij} (\textbf{k})\, , \qquad \mathcal{W}_{ij} (\textbf{k}) \equiv \int_{|\textbf{x}|< \mathcal{R}} \dd^3 x \, \left(3 x_i x_j - |\textbf{x}|^2 \delta_{ij} \right) \ee^{\ii \textbf{k} \cdot \textbf{x}}\, .
    \label{eq:QW_General}
\end{equation}
Considering $x_i = r\,  \hat r_i$ (with $\hat{\textbf{r}}$ a unit radial vector), we also have $\dd^3 x = \dd \Omega \dd r \, r^2$. Writing also $\textbf{k} = k \hat{\textbf{k}}$, with $|\hat{\textbf{k}}| = 1$, we have
\begin{equation}
    \mathcal{W}_{ij} (\textbf{k}) = \int_0^{\mathcal{R}} \dd r \, r^4 \, \int \dd \Omega \left(3 \hat r_i \hat r_j - \delta_{ij} \right) \ee^{\ii k r \, \hat{\textbf{k}} \cdot \hat{\textbf{r}}}\, .
    \label{eq:Wij_1}
\end{equation}
To evaluate $\mathcal{W}_{ij}$, we first notice that its trace is zero, and thus takes the form
\begin{equation}
    \mathcal{W}_{ij} \equiv \mathcal{C}(k) \left[\hat k_i \hat k_j - \frac{\delta_{ij}}{3} \right]\, ,
    \label{eq:Wij_C}
\end{equation}
with $\mathcal{C}(k)$ a scalar function determined by contracting $\mathcal{W}_{ij}$ with $\hat k^i \hat k^j$
\begin{equation}
    \hat k^i \hat k^j \mathcal{W}_{ij} = \frac{2}{3} \mathcal{C}(k)\, .
    \label{eq:CkDef}
\end{equation}
Therefore, contracting \cref{eq:Wij_1} we obtain
\begin{equation}
\begin{split}
    \hat k^i \hat k^j \mathcal{W}_{ij} & = \int_0^{\mathcal{R}} \dd r \, r^4 \int \dd \Omega \left[3 \hat k^i \hat r_i \, \hat k^j \hat r_j -1 \right] \ee^{\ii k r \, \hat{\textbf{k}}\cdot \hat{\textbf{r}}}\\
    & = \int_0^{\mathcal{R}} \dd r \, r^4 \int \dd \Omega \left(3 \cos^2 \theta -1 \right) \ee^{\ii k r \, \cos\theta}\, \\
    & = -\frac{8\pi}{k^5} \left[k \mathcal{R} \left(-15 + k^2 \mathcal{R}^2 \right)\cos k \mathcal{R} + 3 \left(5 - 2k^2 \mathcal{R}^2 \right)\sin k \mathcal{R} \right]\, ,
\end{split}
\end{equation}
where, in the second step, we introduced the angle $\theta$ between $\hat k^i$ and $\hat r^i$. Equating this to \cref{eq:CkDef} gives
\begin{equation}
    \mathcal{C}(k) = -\frac{12\pi}{k^5} \left[k \mathcal{R} \left(-15 + k^2 \mathcal{R}^2 \right)\cos k \mathcal{R} + 3 \left(5 - 2k^2 \mathcal{R}^2 \right)\sin k \mathcal{R} \right]\, . 
\end{equation}
This fully specify $\mathcal{W}_{ij}$ through \cref{eq:Wij_C}.

Going back to \cref{eq:QW_General}, using also \cref{eq:PowerSpectrum_Definition}, we obtain
\begin{equation}
\begin{split}
    \langle Q_{ij} Q^{ij} \rangle &= \tilde \rho_{\text{\tiny FLRW}}^2 \int \frac{\dd^3 k}{(2\pi)^3} P(k) \, \left(\hat k_i \hat k_j - \frac{\delta_{ij}}{3} \right)\left(\hat k^i \hat k^j - \frac{\delta^{ij}}{3} \right) \mathcal{C}(k)^2\\
    &= \frac{2}{3}\tilde \rho_{\text{\tiny FLRW}}^2 \int \frac{\dd^3 k}{(2\pi)^3} P(k) \mathcal{C}(k)^2\\
    & = \frac{\tilde \rho_{\text{\tiny FLRW}}^2}{3\pi^2}  \int_0^\infty \dd k \, k^2 \, P(k) \, \mathcal{C}(k)^2\, .
    \label{eq:QSquared}
\end{split}
\end{equation}
We express the present-day linear matter power spectrum as~\cite{Bardeen:1985tr,Baumann:2022mni} 
\begin{equation}
    P(k, a = 1) = \frac{2\pi^2}{k^3} A_{\rm s} \left(\frac{k}{k_\ast} \right)^{n_{\rm s}-1} \, \left[\frac{2k^2 T(k) D(a = 1)}{5 \Omega_{\rm m} H_0^2} \right]^2\, ,
    \label{eq:PowerSpectrum}
\end{equation}
where $n_{\rm s}$ is the primordial scalar spectral index ($n_{\rm s}\approx 0.96$ according to $Planck$ data~\cite{Planck:2018vyg}), $A_{\rm s}$ is a normalization constant ($A_{\rm s} \sim 10^{-9}$~\cite{Planck:2018vyg}), $k_\ast$ is the pivot scale, here taken as $k_\ast = 5 \cdot 10^{-2} \, \text{Mpc}^{-1}$~\cite{Planck:2018vyg}. $D(a)$ is the linear growth factor describing the growth of the density contrast; during matter domination,  $D(a) \propto a$ for all $k$. Its present-day value in the standard cosmological model is $D(a = 1) \approx 0.761$~\cite{Planck:2018vyg}, which includes the suppression from dark energy. Since this suppression is a feature of the $\Lambda$CDM model and we do not expect the order of magnitude to differ significantly within AFIC, we set $D(a = 1) = 1$ as an order-of-magnitude approximation. A self-consistent evaluation of $D(a)$ in our framework is left to future work. $\Omega_{\rm m}$ and $H_0$ denote the total matter density parameter and the Hubble constant, respectively. Finally, $T(k)$ is the transfer function. We adopt the standard CDM prescription for linear perturbations as a first approximation, leaving a self-consistent determination of the transfer function within the AFIC framework to future work. This approximation is adequate for the order-of-magnitude estimates pursued here. Specifically, we employ the analytical fitting formula derived by Bardeen, Bond, Kaiser and Szalay~\cite{Bardeen:1985tr} (see also Refs.~\cite{1995ApJS..100..281S,Hu:1995en,Eisenstein:1997ik,Amendola:2015ksp})
\begin{equation}
\begin{split}
    T(k) &= \frac{\ln[1+2.34 q(k)]}{2.34 q(k)}\left\{1 + 3.89 q(k) + [16.1 q(k)]^2 + [5.46 q(k)]^3 + [6.71 q(k)]^4 \right\}^{-1/4} \, ,\\
    q(k) &\equiv \frac{k}{\Omega_{\rm m} h^2 \, \text{Mpc}^{-1}} \exp\left[-\Omega_{\rm b} \left(1 + \frac{1}{\Omega_{\rm m}} \right) \right]\, ,
    \label{eq:TransferFunction}
\end{split}
\end{equation}
where $\Omega_{\mathrm b}$ denotes the baryon density parameter. Before evaluating the integral, we need to specify the scale $\mathcal{R}$. As an illustrative estimate, we consider the interval $\mathcal{R} \sim [10^2,\, 10^3]\,\text{Mpc}$. Plugging \cref{eq:PowerSpectrum,eq:TransferFunction} into \cref{eq:QSquared}, evaluating the integral numerically, and using $\tilde\rho_{\text{\tiny FLRW}} \sim \rho_m \sim 4 \cdot 10^{10} M_\odot \text{Mpc}^{-3}$ (consistent with the observed value of the present average matter density) yield the estimate of $\langle Q_{ij} Q^{ij} \rangle$.

Having computed $\langle Q_{ij} Q^{ij} \rangle$, we now turn to the estimate of $Q_{ij} n^i n^j$. Since $Q_{ij}$ is a symmetric traceless tensor, its 
eigenvalues take the form $(\lambda_1, \lambda_2, \lambda_3)$ with 
$\sum_a \lambda_a = 0$. From the tensorial structure~\eqref{eq:Wij_C}, in Fourier space each mode of 
$Q_{ij}$ exhibits the axially symmetric traceless structure
$\hat{k}_i\hat{k}_j - \delta_{ij}/3$, whose eigenvectors along and 
perpendicular to $\mathbf{\hat k}$ have eigenvalues $2/3$ and $-1/3$, 
respectively. Denoting by $n^i$ the dominant eigenvector and writing 
$Q_{ij}n^in^j \equiv 2q$, the eigenvalues thus take the form 
$(2q,\,-q,\,-q)$, so that $ \langle Q_{ij}Q^{ij} \rangle = 6\langle q^2\rangle$, from which, approximately identifying $q \sim \sqrt{\langle q^2 \rangle}$, 
\begin{equation}
    Q_{ij}n^in^j \sim \sqrt{\frac{2}{3}\,\langle Q_{ij}Q^{ij}\rangle}\,.
    \label{eq:Qnn_estimate}
\end{equation}
Since the numerical prefactor $\sqrt{2/3}\approx 0.82$ is of order unity, 
\cref{eq:Qnn_estimate} yields the order-of-magnitude estimate 
$Q_{ij}n^in^j \sim \sqrt{\langle Q_{ij}Q^{ij}\rangle}$. Using the latter together with the numerical estimation of $\langle Q_{ij} Q^{ij}\rangle$ derived above yields the results in \cref{Q}. 
\end{appendix}

\bibliography{refs}

@book{Baumann:2022mni,
    author = "Baumann, Daniel",
    title = "{Cosmology}",
    doi = "10.1017/9781108937092",
    isbn = "978-1-108-93709-2, 978-1-108-83807-8",
    publisher = "Cambridge University Press",
    month = "7",
    year = "2022"
}

@article{deLima:2025bqf,
    author = "de Lima, Leonardo and Rodrigues, Davi C.",
    title = "{Schwarzschild{\textendash}de Sitter spacetime in regular coordinates with cosmological time}",
    eprint = "2508.00951",
    archivePrefix = "arXiv",
    primaryClass = "gr-qc",
    doi = "10.1103/cnx7-ht2l",
    journal = "Phys. Rev. D",
    volume = "112",
    number = "8",
    pages = "084062",
    year = "2025"
}

@article{Riess:1998cb,
    author = "Riess, Adam G. and others",
    collaboration = "High-Z Supernova Search",
    title = "{Observational evidence from supernovae for an accelerating universe and a cosmological constant}",
    eprint = "astro-ph/9805201",
    archivePrefix = "arXiv",
    primaryClass = "astro-ph",
    doi = "10.1086/300499",
    journal = "Astron. J.",
    volume = "116",
    pages = "1009--1038",
    year = "1998"
}

@article{Perlmutter:1998np,
    author = "Perlmutter, S. and others",
    collaboration = "Supernova Cosmology Project",
    title = "{Measurements of Omega and Lambda from 42 High-Z Supernovae}",
    eprint = "astro-ph/9812133",
    archivePrefix = "arXiv",
    primaryClass = "astro-ph",
    doi = "10.1086/307221",
    journal = "Astrophys. J.",
    volume = "517",
    pages = "565--586",
    year = "1999"
}

@article{Perlmutter:1997zf,
    author = "Perlmutter, S. and others",
    collaboration = "Supernova Cosmology Project",
    title = "{Discovery of a supernova explosion at half the age of the Universe and its cosmological implications}",
    eprint = "astro-ph/9712212",
    archivePrefix = "arXiv",
    primaryClass = "astro-ph",
    doi = "10.1038/34124",
    journal = "Nature",
    volume = "391",
    pages = "51--54",
    year = "1998"
}

@article{Carroll:1998zi,
    author = "Carroll, Sean M.",
    title = "{Quintessence and the rest of the world}",
    eprint = "astro-ph/9806099",
    archivePrefix = "arXiv",
    reportNumber = "NSF-ITP-98-063",
    doi = "10.1103/PhysRevLett.81.3067",
    journal = "Phys. Rev. Lett.",
    volume = "81",
    pages = "3067--3070",
    year = "1998"
}

@article{Carroll:1991mt,
    author = "Carroll, Sean M. and Press, William H. and Turner, Edwin L.",
    title = "{The Cosmological constant}",
    reportNumber = "CFA-3332",
    doi = "10.1146/annurev.aa.30.090192.002435",
    journal = "Ann. Rev. Astron. Astrophys.",
    volume = "30",
    pages = "499--542",
    year = "1992"
}

@article{Carroll:2004de,
    author = "Carroll, Sean M. and De Felice, Antonio and Duvvuri, Vikram and Easson, Damien A. and Trodden, Mark and Turner, Michael S.",
    title = "{The Cosmology of generalized modified gravity models}",
    eprint = "astro-ph/0410031",
    archivePrefix = "arXiv",
    reportNumber = "FERMILAB-PUB-04-335-A",
    doi = "10.1103/PhysRevD.71.063513",
    journal = "Phys. Rev. D",
    volume = "71",
    pages = "063513",
    year = "2005"
}

@article{Tsujikawa:2013fta,
    author = "Tsujikawa, Shinji",
    title = "{Quintessence: A Review}",
    eprint = "1304.1961",
    archivePrefix = "arXiv",
    primaryClass = "gr-qc",
    doi = "10.1088/0264-9381/30/21/214003",
    journal = "Class. Quant. Grav.",
    volume = "30",
    pages = "214003",
    year = "2013"
}

@article{DeFelice:2010aj,
    author = "De Felice, Antonio and Tsujikawa, Shinji",
    title = "{f(R) theories}",
    eprint = "1002.4928",
    archivePrefix = "arXiv",
    primaryClass = "gr-qc",
    doi = "10.12942/lrr-2010-3",
    journal = "Living Rev. Rel.",
    volume = "13",
    pages = "3",
    year = "2010"
}

@article{Amendola:2006kh,
    author = "Amendola, Luca and Polarski, David and Tsujikawa, Shinji",
    title = "{Are f(R) dark energy models cosmologically viable ?}",
    eprint = "astro-ph/0603703",
    archivePrefix = "arXiv",
    doi = "10.1103/PhysRevLett.98.131302",
    journal = "Phys. Rev. Lett.",
    volume = "98",
    pages = "131302",
    year = "2007"
}

@article{Amendola:2006we,
    author = "Amendola, Luca and Gannouji, Radouane and Polarski, David and Tsujikawa, Shinji",
    title = "{Conditions for the cosmological viability of f(R) dark energy models}",
    eprint = "gr-qc/0612180",
    archivePrefix = "arXiv",
    doi = "10.1103/PhysRevD.75.083504",
    journal = "Phys. Rev. D",
    volume = "75",
    pages = "083504",
    year = "2007"
}

@article{Davis:1985rj,
    author = "Davis, Marc and Efstathiou, George and Frenk, Carlos S. and White, Simon D. M.",
    title = "{The evolution of large-scale structure in a universe dominated by cold dark matter}",
    doi = "10.1086/163168",
    journal = "Astrophys. J.",
    volume = "292",
    pages = "371--394",
    year = "1985"
}

@article{Efstathiou:1985zz,
    author = "Efstathiou, G. and Davis, M. and White, S. D. M. and Frenk, C. S.",
    title = "{Numerical techniques for large cosmological N-body simulations}",
    doi = "10.1086/191003",
    journal = "Astrophys. J. Suppl.",
    volume = "57",
    pages = "241--260",
    year = "1985"
}

@article{White:1987zz,
    author = "White, Simon D. M. and Frenk, Carlos S. and Davis, Marc and Efstathiou, George",
    title = "{Clustering in a universe dominated by cold dark matter}",
    doi = "10.1086/165005",
    journal = "Astrophys. J.",
    volume = "313",
    pages = "505--516",
    year = "1987"
}

@article{Frenk:1988zz,
    author = "Frenk, Carlos S. and White, Simon D. M. and Davis, Marc and Efstathiou, George",
    title = "{The properties of dark matter halos in a cold dark matter universe}",
    doi = "10.1086/166213",
    journal = "Astrophys. J.",
    volume = "327",
    pages = "507--525",
    year = "1988"
}

@article{KraussTurner1995,
  author  = {Krauss, Lawrence M. and Turner, Michael S.},
  title   = {The cosmological constant is back},
  journal = {General Relativity and Gravitation},
  year    = {1995},
  volume  = {27},
  pages   = {1137--1144},
  doi     = {10.1007/bf02108229}
}

@article{Hellaby:2006cj,
    author = "Hellaby, Charles",
    title = "{The Mass of the Cosmos}",
    eprint = "astro-ph/0603637",
    archivePrefix = "arXiv",
    reportNumber = "UCT-COSMOLOGY-2006-03-23",
    doi = "10.1111/j.1365-2966.2006.10475.x",
    journal = "Mon. Not. Roy. Astron. Soc.",
    volume = "370",
    pages = "239--244",
    year = "2006"
}

@article{Carroll:2000fy,
    author = "Carroll, Sean M.",
    title = "{The Cosmological Constant}",
    eprint = "astro-ph/0004075",
    archivePrefix = "arXiv",
    doi = "10.12942/lrr-2001-1",
    journal = "Living Rev. Rel.",
    volume = "4",
    pages = "1",
    year = "2001"
}

@article{Martin:2012bt,
    author = "Martin, Jerome",
    title = "{Everything You Always Wanted to Know About the Cosmological Constant Problem (But Were Afraid to Ask)}",
    eprint = "1205.3365",
    archivePrefix = "arXiv",
    primaryClass = "astro-ph.CO",
    doi = "10.1016/j.crhy.2012.04.008",
    journal = "C. R. Phys.",
    volume = "13",
    pages = "566--665",
    year = "2012"
}

@article{Zlatev:1998tr,
    author = "Zlatev, Ivaylo and Wang, Limin and Steinhardt, Paul J.",
    title = "{Quintessence, cosmic coincidence, and the cosmological constant}",
    eprint = "astro-ph/9807002",
    archivePrefix = "arXiv",
    doi = "10.1103/PhysRevLett.82.896",
    journal = "Phys. Rev. Lett.",
    volume = "82",
    pages = "896--899",
    year = "1999"
}

@article{Weinberg:1988cp,
    author = "Weinberg, Steven",
    title = "{The Cosmological Constant Problem}",
    doi = "10.1103/RevModPhys.61.1",
    journal = "Rev. Mod. Phys.",
    volume = "61",
    pages = "1--23",
    year = "1989"
}

@article{Granett:2008ju,
    author = "Granett, Benjamin R. and Neyrinck, Mark C. and Szapudi, Istvan",
    title = "{An Imprint of Super-Structures on the Microwave Background due to the Integrated Sachs-Wolfe Effect}",
    eprint = "0805.3695",
    archivePrefix = "arXiv",
    primaryClass = "astro-ph",
    doi = "10.1086/591670",
    journal = "Astrophys. J. Lett.",
    volume = "683",
    pages = "L99--L102",
    year = "2008"
}

@article{Kovacs:2018xqa,
    author = "Kov\'acs, Andr\'as and others",
    collaboration = "DES",
    title = "{More evidence for a large ISW signal from supervoids in the Dark Energy Survey}",
    eprint = "1811.07812",
    archivePrefix = "arXiv",
    primaryClass = "astro-ph.CO",
    doi = "10.1093/mnras/stz172",
    journal = "Mon. Not. Roy. Astron. Soc.",
    volume = "484",
    number = "4",
    pages = "5223--5240",
    year = "2019"
}

@article{DES:2020thz,
    author = "Vielzeuf, P. and others",
    collaboration = "DES",
    title = "{The ISW signal from supervoids in the Dark Energy Survey Year 3 data}",
    eprint = "2004.14814",
    archivePrefix = "arXiv",
    primaryClass = "astro-ph.CO",
    doi = "10.1093/mnras/staa2993",
    journal = "Mon. Not. Roy. Astron. Soc.",
    volume = "500",
    number = "1",
    pages = "464--481",
    year = "2020"
}

@article{Kovacs:2021edg,
    author = "Kov\'acs, Andr\'as and others",
    title = "{The DESI N-body void catalogue and the ISW effect}",
    eprint = "2111.08544",
    archivePrefix = "arXiv",
    primaryClass = "astro-ph.CO",
    doi = "10.1093/mnras/stac228",
    journal = "Mon. Not. Roy. Astron. Soc.",
    volume = "513",
    number = "1",
    pages = "15--32",
    year = "2022"
}

@article{Cadoni:2026cyh,
    author = "Cadoni, Mariano and Cozzumbo, Andrea and Murgia, Riccardo and Pili, Matteo and Pitzalis, Mirko and Sanna, Andrea P. and Schulz, Matteo and Usai, Ilaria",
    title = "{Late-time cosmic acceleration from matter inhomogeneities: constraints from DESI DR2 and SNe Ia}",
    eprint = "2611.31374",
    archivePrefix = "arXiv",
    primaryClass = "astro-ph.CO",
    month = "11",
    year = "2026"
}

@article{Bartelmann:2010fz,
    author = "Bartelmann, Matthias",
    title = "{Gravitational Lensing}",
    eprint = "1010.3829",
    archivePrefix = "arXiv",
    primaryClass = "astro-ph.CO",
    doi = "10.1088/0264-9381/27/23/233001",
    journal = "Class. Quant. Grav.",
    volume = "27",
    pages = "233001",
    year = "2010"
}

@article{Hoekstra:2008db,
    author = "Hoekstra, Henk and Jain, Bhuvnesh",
    title = "{Weak Gravitational Lensing and its Cosmological Applications}",
    eprint = "0805.0139",
    archivePrefix = "arXiv",
    primaryClass = "astro-ph",
    doi = "10.1146/annurev.nucl.58.110707.171151",
    journal = "Ann. Rev. Nucl. Part. Sci.",
    volume = "58",
    pages = "99--123",
    year = "2008"
}

@article{Fu:2007qq,
    author = "Fu, L. and others",
    title = "{Very weak lensing in the CFHTLS Wide: Cosmology from cosmic shear in the linear regime}",
    eprint = "0712.0884",
    archivePrefix = "arXiv",
    primaryClass = "astro-ph",
    doi = "10.1051/0004-6361:20078522",
    journal = "Astron. Astrophys.",
    volume = "479",
    pages = "9--25",
    year = "2008"
}

@article{Munshi:2006fn,
    author = "Munshi, D. and Valageas, P. and Van Waerbeke, Ludovic and Heavens, A.",
    title = "{Cosmology with Weak Lensing Surveys}",
    eprint = "astro-ph/0612667",
    archivePrefix = "arXiv",
    doi = "10.1016/j.physrep.2008.02.003",
    journal = "Phys. Rept.",
    volume = "462",
    pages = "67--121",
    year = "2008"
}

@article{Refregier:2003ct,
    author = "Refregier, Alexandre",
    title = "{Weak gravitational lensing by large scale structure}",
    eprint = "astro-ph/0307212",
    archivePrefix = "arXiv",
    doi = "10.1146/annurev.astro.41.111302.102207",
    journal = "Ann. Rev. Astron. Astrophys.",
    volume = "41",
    pages = "645--668",
    year = "2003"
}

@article{Kumar:2020ole,
    author = "Kumar, Darshan and Jain, Deepak and Mahajan, Shobhit and Mukherjee, Amitabha and Rani, Nisha",
    title = "{Constraining cosmological and galaxy parameters using strong gravitational lensing systems}",
    eprint = "2002.06354",
    archivePrefix = "arXiv",
    primaryClass = "astro-ph.CO",
    doi = "10.1103/PhysRevD.103.063511",
    journal = "Phys. Rev. D",
    volume = "103",
    number = "6",
    pages = "063511",
    year = "2021"
}

@article{Cao:2019kgn,
    author = "Cao, Shuo and Qi, Jingzhao and Cao, Zhoujian and Biesiada, Marek and Li, Jin and Pan, Yu and Zhu, Zong-Hong",
    title = "{Direct test of the FLRW metric from strongly lensed gravitational wave observations}",
    eprint = "1910.10365",
    archivePrefix = "arXiv",
    primaryClass = "astro-ph.CO",
    doi = "10.1038/s41598-019-47616-4",
    journal = "Sci. Rep.",
    volume = "9",
    number = "1",
    pages = "11608",
    year = "2019"
}

@article{Liao:2019hfl,
    author = "Liao, Kai",
    title = "{Constraints on cosmic curvature with lensing time delays and gravitational waves}",
    eprint = "1904.01744",
    archivePrefix = "arXiv",
    primaryClass = "astro-ph.CO",
    doi = "10.1103/PhysRevD.99.083514",
    journal = "Phys. Rev. D",
    volume = "99",
    number = "8",
    pages = "083514",
    year = "2019"
}

@article{Li:2019rns,
    author = "Li, Yufeng and Fan, Xilong and Gou, Lijun",
    title = "{Constraining Cosmological Parameters in the FLRW Metric with Lensed GW+EM Signals}",
    eprint = "1901.10638",
    archivePrefix = "arXiv",
    primaryClass = "astro-ph.CO",
    doi = "10.3847/1538-4357/ab037e",
    journal = "Astrophys. J.",
    volume = "873",
    number = "1",
    pages = "37",
    year = "2019"
}

@article{Wagner:2018jxp,
    author = "Wagner, Jenny and Meyer, Sven",
    title = "{Generalised model-independent characterisation of strong gravitational lenses V: reconstructing the lensing distance ratio by supernovae for a general Friedmann universe}",
    eprint = "1812.04002",
    archivePrefix = "arXiv",
    primaryClass = "astro-ph.CO",
    doi = "10.1093/mnras/stz2717",
    journal = "Mon. Not. Roy. Astron. Soc.",
    volume = "490",
    number = "2",
    pages = "1913--1927",
    year = "2019"
}

@article{Qi:2018aio,
    author = "Qi, Jing-Zhao and Cao, Shuo and Zhang, Sixuan and Biesiada, Marek and Wu, Yan and Zhu, Zong-Hong",
    title = "{The distance sum rule from strong lensing systems and quasars {\textendash} test of cosmic curvature and beyond}",
    eprint = "1803.01990",
    archivePrefix = "arXiv",
    primaryClass = "astro-ph.CO",
    doi = "10.1093/mnras/sty3175",
    journal = "Mon. Not. Roy. Astron. Soc.",
    volume = "483",
    number = "1",
    pages = "1104--1113",
    year = "2019"
}

@article{Qi:2018atg,
    author = "Qi, Jingzhao and Cao, Shuo and Biesiada, Marek and Ding, Xuheng and Zhu, Zong-Hong and Zheng, Xiaogang",
    title = "{Strongly gravitationally lensed type Ia supernovae: Direct test of the Friedman-Lema{\^\i}tre-Robertson-Walker metric}",
    eprint = "1802.05532",
    archivePrefix = "arXiv",
    primaryClass = "astro-ph.CO",
    doi = "10.1103/PhysRevD.100.023530",
    journal = "Phys. Rev. D",
    volume = "100",
    number = "2",
    pages = "023530",
    year = "2019"
}

@article{Liao:2017yeq,
    author = "Liao, Kai and Li, Zhengxiang and Wang, Guo-Jian and Fan, Xi-Long",
    title = "{Test of the FLRW metric and curvature with strong lens time delays}",
    eprint = "1704.04329",
    archivePrefix = "arXiv",
    primaryClass = "astro-ph.CO",
    doi = "10.3847/1538-4357/aa697e",
    journal = "Astrophys. J.",
    volume = "839",
    number = "2",
    pages = "70",
    year = "2017"
}

@article{Rasanen:2014mca,
    author = {R{\"a}s{\"a}nen, Syksy and Bolejko, Krzysztof and Finoguenov, Alexis},
    title = "{New Test of the Friedmann-Lema{\^\i}tre-Robertson-Walker Metric Using the Distance Sum Rule}",
    eprint = "1412.4976",
    archivePrefix = "arXiv",
    primaryClass = "astro-ph.CO",
    reportNumber = "HIP-2014-37-TH",
    doi = "10.1103/PhysRevLett.115.101301",
    journal = "Phys. Rev. Lett.",
    volume = "115",
    number = "10",
    pages = "101301",
    year = "2015"
}

@ARTICLE{shafieloo_clarkson_2010,
       author = {{Shafieloo}, Arman and {Clarkson}, Chris},
        title = "{Model independent tests of the standard cosmological model}",
      journal = {\prd},
         year = 2010,
        month = apr,
       volume = {81},
       number = {8},
          eid = {083537},
        pages = {083537},
          doi = {10.1103/PhysRevD.81.083537},
archivePrefix = {arXiv},
       eprint = {0911.4858},
 primaryClass = {astro-ph.CO},
       adsurl = {https://ui.adsabs.harvard.edu/abs/2010PhRvD..81h3537S}
}

@article{Wambsganss:1998gg,
    author = "Wambsganss, Joachim",
    title = "{Gravitational lensing in astronomy}",
    eprint = "astro-ph/9812021",
    archivePrefix = "arXiv",
    doi = "10.12942/lrr-1998-12",
    journal = "Living Rev. Rel.",
    volume = "1",
    pages = "12",
    year = "1998"
}

@article{Krishnan:2021jmh,
    author = "Krishnan, Chethan and Mohayaee, Roya and Colg{\'a}in, Eoin {\'O}. and Sheikh-Jabbari, M. M. and Yin, L.",
    title = "{Hints of FLRW breakdown from supernovae}",
    eprint = "2106.02532",
    archivePrefix = "arXiv",
    primaryClass = "astro-ph.CO",
    doi = "10.1103/PhysRevD.105.063514",
    journal = "Phys. Rev. D",
    volume = "105",
    number = "6",
    pages = "063514",
    year = "2022"
}

@article{Scrimgeour:2012wt,
    author = "Scrimgeour, Morag and others",
    title = "{The WiggleZ Dark Energy Survey: the transition to large-scale cosmic homogeneity}",
    eprint = "1205.6812",
    archivePrefix = "arXiv",
    primaryClass = "astro-ph.CO",
    doi = "10.1111/j.1365-2966.2012.21402.x",
    journal = "Mon. Not. Roy. Astron. Soc.",
    volume = "425",
    pages = "116--134",
    year = "2012"
}

@article{Wu:1998ad,
    author = "Wu, Kelvin K. S. and Lahav, Ofer and Rees, Martin J.",
    editor = "Sato, H. and Sugiyama, N.",
    title = "{The large-scale smoothness of the Universe}",
    eprint = "astro-ph/9804062",
    archivePrefix = "arXiv",
    reportNumber = "OL-07-98",
    doi = "10.1038/16637",
    journal = "Nature",
    volume = "397",
    pages = "225--230",
    year = "1999"
}

@book{CANTATA:2021asi,
    author = "Akrami, Yashar and others",
    editor = "Saridakis, Emmanuel N. and others",
    collaboration = "CANTATA",
    title = "{Modified Gravity and Cosmology. An Update by the CANTATA Network}",
    eprint = "2105.12582",
    archivePrefix = "arXiv",
    primaryClass = "gr-qc",
    doi = "10.1007/978-3-030-83715-0",
    isbn = "978-3-030-83714-3, 978-3-030-83717-4, 978-3-030-83715-0",
    publisher = "Springer",
    year = "2021"
}

@article{Vagnozzi:2020dfn,
    author = "Vagnozzi, Sunny and Loeb, Abraham and Moresco, Michele",
    title = "{Eppur {\`e} piatto? The Cosmic Chronometers Take on Spatial Curvature and Cosmic Concordance}",
    eprint = "2011.11645",
    archivePrefix = "arXiv",
    primaryClass = "astro-ph.CO",
    doi = "10.3847/1538-4357/abd4df",
    journal = "Astrophys. J.",
    volume = "908",
    number = "1",
    pages = "84",
    year = "2021"
}

@article{Efstathiou:2020wem,
    author = "Efstathiou, George and Gratton, Steven",
    title = "{The evidence for a spatially flat Universe}",
    eprint = "2002.06892",
    archivePrefix = "arXiv",
    primaryClass = "astro-ph.CO",
    doi = "10.1093/mnrasl/slaa093",
    journal = "Mon. Not. Roy. Astron. Soc.",
    volume = "496",
    number = "1",
    pages = "L91--L95",
    year = "2020"
}

@article{Salucci:2018hqu,
    author = "Salucci, Paolo",
    title = "{The distribution of dark matter in galaxies}",
    eprint = "1811.08843",
    archivePrefix = "arXiv",
    primaryClass = "astro-ph.GA",
    doi = "10.1007/s00159-018-0113-1",
    journal = "Astron. Astrophys. Rev.",
    volume = "27",
    number = "1",
    pages = "2",
    year = "2019"
}

@article{Lelli:2016zqa,
    author = "Lelli, Federico and McGaugh, Stacy S. and Schombert, James M.",
    title = "{SPARC: Mass Models for 175 Disk Galaxies with Spitzer Photometry and Accurate Rotation Curves}",
    eprint = "1606.09251",
    archivePrefix = "arXiv",
    primaryClass = "astro-ph.GA",
    doi = "10.3847/0004-6256/152/6/157",
    journal = "Astron. J.",
    volume = "152",
    pages = "157",
    year = "2016"
}

@article{Lelli:2016cui,
    author = "Lelli, Federico and McGaugh, Stacy S. and Schombert, James M. and Pawlowski, Marcel S.",
    title = "{One Law to Rule Them All: The Radial Acceleration Relation of Galaxies}",
    eprint = "1610.08981",
    archivePrefix = "arXiv",
    primaryClass = "astro-ph.GA",
    doi = "10.3847/1538-4357/836/2/152",
    journal = "Astrophys. J.",
    volume = "836",
    number = "2",
    pages = "152",
    year = "2017"
}

@article{Strigari:2012acq,
    author = "Strigari, Louis E.",
    title = "{Galactic Searches for Dark Matter}",
    eprint = "1211.7090",
    archivePrefix = "arXiv",
    primaryClass = "astro-ph.CO",
    doi = "10.1016/j.physrep.2013.05.004",
    journal = "Phys. Rept.",
    volume = "531",
    pages = "1--88",
    year = "2013"
}

@article{Frenk:2012ph,
    author = "Frenk, C. S. and White, Simon D. M.",
    title = "{Dark matter and cosmic structure}",
    eprint = "1210.0544",
    archivePrefix = "arXiv",
    primaryClass = "astro-ph.CO",
    doi = "10.1002/andp.201200212",
    journal = "Annalen Phys.",
    volume = "524",
    pages = "507--534",
    year = "2012"
}

@article{Giusti:2019wdx,
    author = "Giusti, Andrea",
    title = "{On the corpuscular theory of gravity}",
    doi = "10.1142/S0219887819300010",
    journal = "Int. J. Geom. Meth. Mod. Phys.",
    volume = "16",
    number = "03",
    pages = "1930001",
    year = "2019"
}

@article{Mukhanov:1990me,
    author = "Mukhanov, Viatcheslav F. and Feldman, H. A. and Brandenberger, Robert H.",
    title = "{Theory of cosmological perturbations. Part 1. Classical perturbations. Part 2. Quantum theory of perturbations. Part 3. Extensions}",
    reportNumber = "BROWN-HET-796, BROWN-HET-800, BROWN-HET-780",
    doi = "10.1016/0370-1573(92)90044-Z",
    journal = "Phys. Rept.",
    volume = "215",
    pages = "203--333",
    year = "1992"
}

@article{Galoppo:2025rkd,
    author = "Galoppo, Marco and Mourier, Pierre",
    title = "{Cosmological perturbations on an averaged background}",
    eprint = "2511.17160",
    archivePrefix = "arXiv",
    primaryClass = "gr-qc",
    doi = "10.1088/1475-7516/2026/02/066",
    journal = "JCAP",
    volume = "02",
    pages = "066",
    year = "2026"
}

@article{Racz:2016rss,
    author = "R{\'a}cz, G{\'a}bor and Dobos, L{\'a}szl{\'o} and Beck, R{\'o}bert and Szapudi, Istv{\'a}n and Csabai, Istv{\'a}n",
    title = "{Concordance cosmology without dark energy}",
    eprint = "1607.08797",
    archivePrefix = "arXiv",
    primaryClass = "astro-ph.CO",
    doi = "10.1093/mnrasl/slx026",
    journal = "Mon. Not. Roy. Astron. Soc.",
    volume = "469",
    number = "1",
    pages = "L1--L5",
    year = "2017"
}

@article{CosmoVerseNetwork:2025alb,
    author = "Di Valentino, Eleonora and others",
    collaboration = "CosmoVerse Network",
    title = "{The CosmoVerse White Paper: Addressing observational tensions in cosmology with systematics and fundamental physics}",
    eprint = "2504.01669",
    archivePrefix = "arXiv",
    primaryClass = "astro-ph.CO",
    doi = "10.1016/j.dark.2025.101965",
    journal = "Phys. Dark Univ.",
    volume = "49",
    pages = "101965",
    year = "2025"
}

@article{Scherer:2025esj,
    author = "Scherer, Mateus and Sabogal, Miguel A. and Nunes, Rafael C. and De Felice, Antonio",
    title = "{Challenging the {\ensuremath{\Lambda}}CDM model: 5{\ensuremath{\sigma}} evidence for a dynamical dark energy late-time transition}",
    eprint = "2504.20664",
    archivePrefix = "arXiv",
    primaryClass = "astro-ph.CO",
    doi = "10.1103/n86r-sjgm",
    journal = "Phys. Rev. D",
    volume = "112",
    number = "4",
    pages = "043513",
    year = "2025"
}

@article{Peebles:2002gy,
    author = "Peebles, P. J. E. and Ratra, Bharat",
    editor = "Hsu, Jong-Ping and Fine, D.",
    title = "{The Cosmological Constant and Dark Energy}",
    eprint = "astro-ph/0207347",
    archivePrefix = "arXiv",
    reportNumber = "KSUPT-02-3",
    doi = "10.1103/RevModPhys.75.559",
    journal = "Rev. Mod. Phys.",
    volume = "75",
    pages = "559--606",
    year = "2003"
}

@article{Copeland:2006wr,
    author = "Copeland, Edmund J. and Sami, M. and Tsujikawa, Shinji",
    title = "{Dynamics of dark energy}",
    eprint = "hep-th/0603057",
    archivePrefix = "arXiv",
    doi = "10.1142/S021827180600942X",
    journal = "Int. J. Mod. Phys. D",
    volume = "15",
    pages = "1753--1936",
    year = "2006"
}

@article{Koksbang:2026jxb,
    author = "Koksbang, S. M. and Heinesen, A.",
    title = "{Diagnostic Consistency Tests of the Concordance Cosmology}",
    eprint = "2604.05836",
    archivePrefix = "arXiv",
    primaryClass = "astro-ph.CO",
    doi = "10.1103/x2m7-dtgd",
    journal = "Phys. Rev. Lett.",
    volume = "137",
    number = "9",
    pages = "091001",
    year = "2026"
}

@article{Koksbang:2026wvh,
    author = "Koksbang, S. M. and Heinesen, A.",
    title = "{Model-independent constraints on generalized FLRW consistency relations with bootstrap-based symbolic regression}",
    eprint = "2604.05822",
    archivePrefix = "arXiv",
    primaryClass = "astro-ph.CO",
    doi = "10.1103/jy1h-5zt1",
    journal = "Phys. Rev. D",
    volume = "114",
    number = "4",
    pages = "043549",
    year = "2026"
}

@article{Heinesen:2026wwy,
    author = "Heinesen, Asta and Clifton, Timothy",
    title = "{Observational tests for distinguishing classes of cosmological models}",
    eprint = "2604.07244",
    archivePrefix = "arXiv",
    primaryClass = "astro-ph.CO",
    doi = "10.1103/nrt8-bgnf",
    journal = "Phys. Rev. D",
    volume = "114",
    number = "8",
    pages = "083510",
    year = "2026"
}

@article{Courtois:2025xcs,
    author = "Courtois, Helene M. and Mould, Jeremy and Hollinger, Amber M. and Dupuy, Alexandra and Zhang, Chuan-Peng",
    title = "{In search of the Local Universe dynamical homogeneity scale with CF4++ peculiar velocities}",
    eprint = "2502.01308",
    archivePrefix = "arXiv",
    primaryClass = "astro-ph.CO",
    doi = "10.1051/0004-6361/202553677",
    journal = "Astron. Astrophys.",
    volume = "701",
    pages = "A187",
    year = "2025"
}

@article{Galoppo:2026rin,
    author = "Galoppo, Marco and Buchert, Thomas and Mourier, Pierre",
    title = "{Backreaction and the Role of Spatial Curvature in the Cosmic Neighborhood}",
    eprint = "2605.05454",
    archivePrefix = "arXiv",
    primaryClass = "astro-ph.CO",
    doi = "10.3847/2041-8213/ae5f73",
    journal = "Astrophys. J. Lett.",
    volume = "1002",
    number = "2",
    pages = "L39",
    year = "2026"
}

@article{Linder:2005in,
    author = "Linder, Eric V.",
    title = "{Cosmic growth history and expansion history}",
    eprint = "astro-ph/0507263",
    archivePrefix = "arXiv",
    doi = "10.1103/PhysRevD.72.043529",
    journal = "Phys. Rev. D",
    volume = "72",
    pages = "043529",
    year = "2005"
}

@article{Brax:2017idh,
    author = "Brax, Philippe",
    title = "{What makes the Universe accelerate? A review on what dark energy could be and how to test it}",
    doi = "10.1088/1361-6633/aa8e64",
    journal = "Rept. Prog. Phys.",
    volume = "81",
    number = "1",
    pages = "016902",
    year = "2018"
}

@article{Bengaly:2017slg,
    author = "Bengaly, C. A. P. and Maartens, R. and Santos, M. G.",
    title = "{Probing the Cosmological Principle in the counts of radio galaxies at different frequencies}",
    eprint = "1710.08804",
    archivePrefix = "arXiv",
    primaryClass = "astro-ph.CO",
    doi = "10.1088/1475-7516/2018/04/031",
    journal = "JCAP",
    volume = "04",
    pages = "031",
    year = "2018"
}

@article{Bengaly:2018ykb,
    author = "Bengaly, Carlos A. P. and Siewert, Thilo M. and Schwarz, Dominik J. and Maartens, Roy",
    title = "{Testing the standard model of cosmology with the SKA: the cosmic radio dipole}",
    eprint = "1810.04960",
    archivePrefix = "arXiv",
    primaryClass = "astro-ph.CO",
    doi = "10.1093/mnras/stz832",
    journal = "Mon. Not. Roy. Astron. Soc.",
    volume = "486",
    number = "1",
    pages = "1350--1357",
    year = "2019"
}

@article{Bengaly:2016amk,
    author = "Bengaly, C. A. P. and Bernui, A. and Alcaniz, J. S. and Xavier, H. S. and Novaes, C. P.",
    title = "{Is there evidence for anomalous dipole anisotropy in the large-scale structure?}",
    eprint = "1606.06751",
    archivePrefix = "arXiv",
    primaryClass = "astro-ph.CO",
    doi = "10.1093/mnras/stw2268",
    journal = "Mon. Not. Roy. Astron. Soc.",
    volume = "464",
    number = "1",
    pages = "768--774",
    year = "2017"
}

@article{Blomqvist:2008ud,
    author = "Blomqvist, Michael and Mortsell, Edvard and Nobili, Serena",
    title = "{Probing Dark Energy Inhomogeneities with Supernovae}",
    eprint = "0806.0496",
    archivePrefix = "arXiv",
    primaryClass = "astro-ph",
    doi = "10.1088/1475-7516/2008/06/027",
    journal = "JCAP",
    volume = "06",
    pages = "027",
    year = "2008"
}

@article{Colin:2010ds,
    author = "Colin, Jacques and Mohayaee, Roya and Sarkar, Subir and Shafieloo, Arman",
    title = "{Probing the anisotropic local universe and beyond with SNe Ia data}",
    eprint = "1011.6292",
    archivePrefix = "arXiv",
    primaryClass = "astro-ph.CO",
    doi = "10.1111/j.1365-2966.2011.18402.x",
    journal = "Mon. Not. Roy. Astron. Soc.",
    volume = "414",
    pages = "264--271",
    year = "2011"
}

@article{Alonso:2014xca,
    author = "Alonso, David and Salvador, Ana Isabel and S{\'a}nchez, Francisco Javier and Bilicki, Maciej and Garc{\'\i}a-Bellido, Juan and S{\'a}nchez, Eusebio",
    title = "{Homogeneity and isotropy in the Two Micron All Sky Survey Photometric Redshift catalogue}",
    eprint = "1412.5151",
    archivePrefix = "arXiv",
    primaryClass = "astro-ph.CO",
    doi = "10.1093/mnras/stv309",
    journal = "Mon. Not. Roy. Astron. Soc.",
    volume = "449",
    number = "1",
    pages = "670--684",
    year = "2015"
}

@article{Secrest:2020has,
    author = "Secrest, Nathan J. and von Hausegger, Sebastian and Rameez, Mohamed and Mohayaee, Roya and Sarkar, Subir and Colin, Jacques",
    title = "{A Test of the Cosmological Principle with Quasars}",
    eprint = "2009.14826",
    archivePrefix = "arXiv",
    primaryClass = "astro-ph.CO",
    doi = "10.3847/2041-8213/abdd40",
    journal = "Astrophys. J. Lett.",
    volume = "908",
    number = "2",
    pages = "L51",
    year = "2021"
}

@article{Camarena:2025upt,
    author = "Camarena, David and Greene, Kylar and Houghteling, John and Cyr-Racine, Francis-Yan",
    title = "{Designing concordant distances in the age of precision cosmology: The impact of density fluctuations}",
    eprint = "2507.17969",
    archivePrefix = "arXiv",
    primaryClass = "astro-ph.CO",
    doi = "10.1103/1gj5-3x4m",
    journal = "Phys. Rev. D",
    volume = "112",
    number = "8",
    pages = "083526",
    year = "2025"
}

@article{Adamek:2024hme,
    author = "Adamek, Julian and Clarkson, Chris and Durrer, Ruth and Heinesen, Asta and Kunz, Martin and Macpherson, Hayley J.",
    title = "{Towards Cosmography of the Local Universe}",
    eprint = "2402.12165",
    archivePrefix = "arXiv",
    primaryClass = "astro-ph.CO",
    doi = "10.33232/001c.118782",
    journal = "Open J. Astrophys.",
    volume = "7",
    pages = "001c.118782",
    year = "2024"
}

@article{Camarena:2022iae,
    author = "Camarena, David and Marra, Valerio and Sakr, Ziad and Clarkson, Chris",
    title = "{A void in the Hubble tension? The end of the line for the Hubble bubble}",
    eprint = "2205.05422",
    archivePrefix = "arXiv",
    primaryClass = "astro-ph.CO",
    doi = "10.1088/1361-6382/ac8635",
    journal = "Class. Quant. Grav.",
    volume = "39",
    number = "18",
    pages = "184001",
    year = "2022"
}

@article{Macpherson:2022eve,
    author = "Macpherson, Hayley J.",
    title = "{Cosmological distances with general-relativistic ray tracing: framework and comparison to cosmographic predictions}",
    eprint = "2209.06775",
    archivePrefix = "arXiv",
    primaryClass = "astro-ph.CO",
    doi = "10.1088/1475-7516/2023/03/019",
    journal = "JCAP",
    volume = "03",
    pages = "019",
    year = "2023"
}

@article{Macpherson:2021gbh,
    author = "Macpherson, Hayley J. and Heinesen, Asta",
    title = "{Luminosity distance and anisotropic sky-sampling at low redshifts: A numerical relativity study}",
    eprint = "2103.11918",
    archivePrefix = "arXiv",
    primaryClass = "astro-ph.CO",
    doi = "10.1103/PhysRevD.104.023525",
    journal = "Phys. Rev. D",
    volume = "104",
    pages = "023525",
    year = "2021",
    note = "[Erratum: Phys.Rev.D 104, 109901 (2021)]"
}

@article{Bolejko:2015gmk,
    author = "Bolejko, Krzysztof and Nazer, M. Ahsan and Wiltshire, David L.",
    title = "{Differential cosmic expansion and the Hubble flow anisotropy}",
    eprint = "1512.07364",
    archivePrefix = "arXiv",
    primaryClass = "astro-ph.CO",
    doi = "10.1088/1475-7516/2016/06/035",
    journal = "JCAP",
    volume = "06",
    pages = "035",
    year = "2016"
}

@article{Giani:2021gbs,
    author = "Giani, Leonardo and Piattella, Oliver F. and Kamenshchik, Alexander Yu.",
    title = "{Bianchi IX gravitational collapse of matter inhomogeneities}",
    eprint = "2112.01869",
    archivePrefix = "arXiv",
    primaryClass = "gr-qc",
    doi = "10.1088/1475-7516/2022/03/028",
    journal = "JCAP",
    volume = "03",
    number = "03",
    pages = "028",
    year = "2022"
}

@article{Labini:2025dnc,
    author = "Labini, Francesco Sylos and Antal, Tibor",
    title = "{Large-scale galaxy correlations from the DESI first data release}",
    eprint = "2511.21585",
    archivePrefix = "arXiv",
    primaryClass = "astro-ph.CO",
    doi = "10.1051/0004-6361/202558271",
    journal = "Astron. Astrophys.",
    volume = "707",
    pages = "A254",
    year = "2026"
}

@article{Dias:2023qrl,
    author = "Dias, Bruno L. and Avila, Felipe and Bernui, Armando",
    title = "{Probing cosmic homogeneity in the Local Universe}",
    eprint = "2310.04594",
    archivePrefix = "arXiv",
    primaryClass = "astro-ph.CO",
    doi = "10.1093/mnras/stad2980",
    journal = "Mon. Not. Roy. Astron. Soc.",
    volume = "526",
    number = "3",
    pages = "3219--3229",
    year = "2023"
}

@article{Pandey:2015xea,
    author = "Pandey, Biswajit and Sarkar, Suman",
    title = "{Testing homogeneity in the Sloan Digital Sky Survey Data Release Twelve with Shannon entropy}",
    eprint = "1507.03124",
    archivePrefix = "arXiv",
    primaryClass = "astro-ph.CO",
    doi = "10.1093/mnras/stv2166",
    journal = "Mon. Not. Roy. Astron. Soc.",
    volume = "454",
    number = "3",
    pages = "2647--2656",
    year = "2015"
}

@article{Hogg:2004vw,
    author = "Hogg, David W. and Eisenstein, Daniel J. and Blanton, Michael R. and Bahcall, Neta A. and Brinkmann, J. and Gunn, James E. and Schneider, Donald P.",
    title = "{Cosmic homogeneity demonstrated with luminous red galaxies}",
    eprint = "astro-ph/0411197",
    archivePrefix = "arXiv",
    doi = "10.1086/429084",
    journal = "Astrophys. J.",
    volume = "624",
    pages = "54--58",
    year = "2005"
}

@article{Celerier:2024dvs,
    author = {C{\'e}l{\'e}rier, Marie-No{\"e}lle},
    title = "{Precision cosmology with exact inhomogeneous solutions of general relativity: The Szekeres models}",
    eprint = "2407.04452",
    archivePrefix = "arXiv",
    primaryClass = "gr-qc",
    doi = "10.1103/PhysRevD.110.123526",
    journal = "Phys. Rev. D",
    volume = "110",
    number = "12",
    pages = "123526",
    year = "2024"
}

@article{Szekeres:1975dx,
    author = "Szekeres, Peter",
    title = "{Quasispherical Gravitational Collapse}",
    reportNumber = "PRINT-75-0475 (ADELAIDE)",
    doi = "10.1103/PhysRevD.12.2941",
    journal = "Phys. Rev. D",
    volume = "12",
    pages = "2941",
    year = "1975"
}

@article{Lapi:2023plb,
    author = "Lapi, Andrea and Boco, Lumen and Cueli, Marcos M. and Haridasu, Balakrishna S. and Ronconi, Tommaso and Baccigalupi, Carlo and Danese, Luigi",
    title = "{Little Ado about Everything: {\ensuremath{\eta}}CDM, a Cosmological Model with Fluctuation-driven Acceleration at Late Times}",
    eprint = "2310.06028",
    archivePrefix = "arXiv",
    primaryClass = "astro-ph.CO",
    doi = "10.3847/1538-4357/ad01bb",
    journal = "Astrophys. J.",
    volume = "959",
    number = "2",
    pages = "83",
    year = "2023"
}

@article{Giani:2024nnv,
    author = "Giani, Leonardo and Von Marttens, Rodrigo and Camilleri, Ryan",
    title = "{Novel Approach to Cosmological Nonlinearities as an Effective Fluid}",
    eprint = "2410.15295",
    archivePrefix = "arXiv",
    primaryClass = "astro-ph.CO",
    doi = "10.1103/zr92-m7py",
    journal = "Phys. Rev. Lett.",
    volume = "135",
    number = "7",
    pages = "071004",
    year = "2025"
}

@article{Seifert:2024bqr,
    author = "Seifert, Antonia and Lane, Zachary G. and Galoppo, Marco and Ridden-Harper, Ryan and Wiltshire, David L.",
    title = "{Supernovae evidence for foundational change to cosmological models}",
    eprint = "2412.15143",
    archivePrefix = "arXiv",
    primaryClass = "astro-ph.CO",
    doi = "10.1093/mnrasl/slae112",
    journal = "Mon. Not. Roy. Astron. Soc.",
    volume = "537",
    number = "1",
    pages = "L55--L60",
    year = "2025"
}

@article{Camarena:2021mjr,
    author = "Camarena, David and Marra, Valerio and Sakr, Ziad and Clarkson, Chris",
    title = "{The Copernican principle in light of the latest cosmological~data}",
    eprint = "2107.02296",
    archivePrefix = "arXiv",
    primaryClass = "astro-ph.CO",
    doi = "10.1093/mnras/stab3077",
    journal = "Mon. Not. Roy. Astron. Soc.",
    volume = "509",
    number = "1",
    pages = "1291--1302",
    year = "2021"
}

@article{Oestreicher:2024mtw,
    author = "Oestreicher, Alexander and Koksbang, Sofie Marie",
    title = "{Backreaction in Numerical Relativity: Averaging on Newtonian gauge-like hypersurfaces in Einstein Toolkit cosmological simulations}",
    eprint = "2408.03049",
    archivePrefix = "arXiv",
    primaryClass = "astro-ph.CO",
    doi = "10.33232/001c.124744",
    journal = "Open J. Astrophys.",
    volume = "7",
    pages = "2024",
    month = "8",
    year = "2024"
}

@article{Macpherson:2018btl,
    author = "Macpherson, Hayley J. and Price, Daniel J. and Lasky, Paul D.",
    title = "{Einstein{\textquoteright}s Universe: Cosmological structure formation in numerical relativity}",
    eprint = "1807.01711",
    archivePrefix = "arXiv",
    primaryClass = "astro-ph.CO",
    doi = "10.1103/PhysRevD.99.063522",
    journal = "Phys. Rev. D",
    volume = "99",
    number = "6",
    pages = "063522",
    year = "2019"
}

@article{Breton:2020puw,
    author = "Breton, Michel-Andr{\`e}s and Fleury, Pierre",
    title = "{Theoretical and numerical perspectives on cosmic distance averages}",
    eprint = "2012.07802",
    archivePrefix = "arXiv",
    primaryClass = "astro-ph.CO",
    reportNumber = "IFT-UAM/CSIC-20-183",
    doi = "10.1051/0004-6361/202040140",
    journal = "Astron. Astrophys.",
    volume = "655",
    pages = "A54",
    year = "2021"
}

@article{Brunswic:2020bjx,
    author = "Brunswic, L{\'e}o and Buchert, Thomas",
    title = "{Gauss{\textendash}Bonnet{\textendash}Chern approach to the averaged Universe}",
    eprint = "2002.08336",
    archivePrefix = "arXiv",
    primaryClass = "gr-qc",
    doi = "10.1088/1361-6382/abae45",
    journal = "Class. Quant. Grav.",
    volume = "37",
    number = "21",
    pages = "215022",
    year = "2020"
}

@article{DESI:2025fii,
    author = "Lodha, K. and others",
    collaboration = "DESI",
    title = "{Extended dark energy analysis using DESI DR2 BAO measurements}",
    eprint = "2503.14743",
    archivePrefix = "arXiv",
    primaryClass = "astro-ph.CO",
    reportNumber = "FERMILAB-PUB-25-0164-PPD",
    doi = "10.1103/w4c6-1r5j",
    journal = "Phys. Rev. D",
    volume = "112",
    number = "8",
    pages = "083511",
    year = "2025"
}

@article{DESI:2024aqx,
    author = "Calderon, R. and others",
    collaboration = "DESI",
    title = "{DESI 2024: reconstructing dark energy using crossing statistics with DESI DR1 BAO data}",
    eprint = "2405.04216",
    archivePrefix = "arXiv",
    primaryClass = "astro-ph.CO",
    doi = "10.1088/1475-7516/2024/10/048",
    journal = "JCAP",
    volume = "10",
    pages = "048",
    year = "2024"
}

@article{Galoppo:2025hzy,
    author = "Galoppo, Marco and Giani, Leonardo and Hills, Morag and Valade, Aur{\'e}lien",
    title = "{An effective {\ensuremath{\Lambda}}-Szekeres modelling of the local Universe with Cosmicflows-4}",
    eprint = "2512.16591",
    archivePrefix = "arXiv",
    primaryClass = "astro-ph.CO",
    doi = "10.1088/1475-7516/2026/06/096",
    journal = "JCAP",
    volume = "06",
    pages = "096",
    year = "2026"
}

@article{Marra:2022ixf,
    author = "Marra, V. and Castro, T. and Camarena, D. and Borgani, S. and Ragagnin, A.",
    title = "{The BEHOMO project: {\ensuremath{\Lambda}} Lema{\^\i}tre-Tolman-Bondi N-body simulations}",
    eprint = "2203.04009",
    archivePrefix = "arXiv",
    primaryClass = "astro-ph.CO",
    doi = "10.1051/0004-6361/202243539",
    journal = "Astron. Astrophys.",
    volume = "664",
    pages = "A179",
    year = "2022"
}

@article{Koksbang:2020qry,
    author = "Koksbang, S. M.",
    title = "{On the relationship between mean observations, spatial averages and the Dyer-Roeder approximation in Einstein-Straus models}",
    eprint = "2010.04500",
    archivePrefix = "arXiv",
    primaryClass = "astro-ph.CO",
    doi = "10.1088/1475-7516/2020/11/061",
    journal = "JCAP",
    volume = "11",
    pages = "061",
    year = "2020"
}

@article{Koksbang:2020zej,
    author = "Koksbang, S. M.",
    title = "{Observations in statistically homogeneous, locally inhomogeneous cosmological toy-models without FLRW backgrounds}",
    eprint = "2008.07108",
    archivePrefix = "arXiv",
    primaryClass = "astro-ph.CO",
    reportNumber = "HIP-2020-25/TH",
    doi = "10.1093/mnrasl/slaa146",
    journal = "Mon. Not. Roy. Astron. Soc.",
    volume = "498",
    number = "1",
    pages = "L135--L139",
    year = "2020",
    note = "[Erratum: Mon.Not.Roy.Astron.Soc. 500, (2021)]"
}

@article{Koksbang:2019cen,
    author = "Koksbang, S. M.",
    title = "{Towards statistically homogeneous and isotropic perfect fluid universes with cosmic backreaction}",
    eprint = "1907.08681",
    archivePrefix = "arXiv",
    primaryClass = "gr-qc",
    reportNumber = "HIP-2019-25/TH",
    doi = "10.1088/1361-6382/ab376c",
    journal = "Class. Quant. Grav.",
    volume = "36",
    number = "18",
    pages = "185004",
    year = "2019"
}

@article{Koksbang:2017arw,
    author = "Koksbang, S. M.",
    title = "{Light propagation in Swiss cheese models of random close-packed Szekeres structures: Effects of anisotropy and comparisons with perturbative results}",
    eprint = "1703.03572",
    archivePrefix = "arXiv",
    primaryClass = "astro-ph.CO",
    doi = "10.1103/PhysRevD.95.063532",
    journal = "Phys. Rev. D",
    volume = "95",
    number = "6",
    pages = "063532",
    year = "2017"
}

@article{Sussman:2011na,
    author = "Sussman, Roberto A.",
    title = "{Back-reaction and effective acceleration in generic LTB dust models}",
    eprint = "1102.2663",
    archivePrefix = "arXiv",
    primaryClass = "gr-qc",
    doi = "10.1088/0264-9381/28/23/235002",
    journal = "Class. Quant. Grav.",
    volume = "28",
    pages = "235002",
    year = "2011"
}

@article{Bolejko:2010eb,
    author = "Bolejko, Krzysztof and Celerier, Marie-Noelle",
    title = "{Szekeres Swiss-Cheese model and supernova observations}",
    eprint = "1005.2584",
    archivePrefix = "arXiv",
    primaryClass = "astro-ph.CO",
    doi = "10.1103/PhysRevD.82.103510",
    journal = "Phys. Rev. D",
    volume = "82",
    pages = "103510",
    year = "2010"
}

@article{Clifton:2019cep,
    author = "Clifton, Timothy and Sussman, Roberto A.",
    title = "{Cosmological Backreaction in Spherical and Plane Symmetric Dust-Filled Space-Times}",
    eprint = "1904.02557",
    archivePrefix = "arXiv",
    primaryClass = "gr-qc",
    doi = "10.1088/1361-6382/ab3a14",
    journal = "Class. Quant. Grav.",
    volume = "36",
    number = "20",
    pages = "205004",
    year = "2019"
}

@article{Clarkson:2011uk,
    author = "Clarkson, Chris and Umeh, Obinna",
    title = "{Is backreaction really small within concordance cosmology?}",
    eprint = "1105.1886",
    archivePrefix = "arXiv",
    primaryClass = "astro-ph.CO",
    doi = "10.1088/0264-9381/28/16/164010",
    journal = "Class. Quant. Grav.",
    volume = "28",
    pages = "164010",
    year = "2011"
}

@article{Hellaby:2012zz,
    author = "Hellaby, Charles",
    title = "{A new type of exact arbitrarily inhomogeneous cosmology: Evolution of deceleration in the flat homogeneous-on-average case}",
    eprint = "1203.3652",
    archivePrefix = "arXiv",
    primaryClass = "gr-qc",
    reportNumber = "UCT-COSMOLOGY-2012-03-16",
    doi = "10.1088/1475-7516/2012/01/043",
    journal = "JCAP",
    volume = "01",
    pages = "043",
    year = "2012"
}

@article{Najera:2020jdm,
    author = "N{\'a}jera, S. and Sussman, R. A.",
    title = "{Pancakes as opposed to Swiss Cheese}",
    eprint = "2010.04027",
    archivePrefix = "arXiv",
    primaryClass = "gr-qc",
    doi = "10.1088/1361-6382/abcaec",
    journal = "Class. Quant. Grav.",
    volume = "38",
    number = "1",
    pages = "015016",
    year = "2021"
}

@article{Biswas:2007gi,
    author = "Biswas, Tirthabir and Notari, Alessio",
    title = "{Swiss-Cheese Inhomogeneous Cosmology and the Dark Energy Problem}",
    eprint = "astro-ph/0702555",
    archivePrefix = "arXiv",
    reportNumber = "IGPG-07-2-2",
    doi = "10.1088/1475-7516/2008/06/021",
    journal = "JCAP",
    volume = "06",
    pages = "021",
    year = "2008"
}

@article{Marra:2007pm,
    author = "Marra, Valerio and Kolb, Edward W. and Matarrese, Sabino and Riotto, Antonio",
    title = "{On cosmological observables in a swiss-cheese universe}",
    eprint = "0708.3622",
    archivePrefix = "arXiv",
    primaryClass = "astro-ph",
    doi = "10.1103/PhysRevD.76.123004",
    journal = "Phys. Rev. D",
    volume = "76",
    pages = "123004",
    year = "2007"
}

@article{Schander:2021pgt,
    author = "Schander, S. and Thiemann, T.",
    title = "{Backreaction in Cosmology}",
    eprint = "2106.06043",
    archivePrefix = "arXiv",
    primaryClass = "gr-qc",
    doi = "10.3389/fspas.2021.692198",
    journal = "Front. Astron. Space Sci.",
    volume = "0",
    pages = "113",
    year = "2021"
}

@article{Green:2013yua,
    author = "Green, Stephen R. and Wald, Robert M.",
    title = "{Examples of backreaction of small scale inhomogeneities in cosmology}",
    eprint = "1304.2318",
    archivePrefix = "arXiv",
    primaryClass = "gr-qc",
    doi = "10.1103/PhysRevD.87.124037",
    journal = "Phys. Rev. D",
    volume = "87",
    number = "12",
    pages = "124037",
    year = "2013"
}

@article{Green:2010qy,
    author = "Green, Stephen R. and Wald, Robert M.",
    title = "{A new framework for analyzing the effects of small scale inhomogeneities in cosmology}",
    eprint = "1011.4920",
    archivePrefix = "arXiv",
    primaryClass = "gr-qc",
    doi = "10.1103/PhysRevD.83.084020",
    journal = "Phys. Rev. D",
    volume = "83",
    pages = "084020",
    year = "2011"
}

@article{Gomes:2024iyq,
    author = "Gomes, Leandro G.",
    title = "{Breaking the cosmological principle into pieces: a prelude to the intrinsically homogeneous and isotropic spacetimes}",
    eprint = "2401.01992",
    archivePrefix = "arXiv",
    primaryClass = "gr-qc",
    doi = "10.1088/1361-6382/ad3609",
    journal = "Class. Quant. Grav.",
    volume = "41",
    number = "9",
    pages = "095004",
    year = "2024"
}

@article{Fanizza:2019pfp,
    author = "Fanizza, Giuseppe and Gasperini, Maurizio and Marozzi, Giovanni and Veneziano, Gabriele",
    title = "{Generalized covariant prescriptions for averaging cosmological observables}",
    eprint = "1911.09469",
    archivePrefix = "arXiv",
    primaryClass = "gr-qc",
    reportNumber = "BA-TH/721-19, CERN-TH-2019-183",
    doi = "10.1088/1475-7516/2020/02/017",
    journal = "JCAP",
    volume = "02",
    pages = "017",
    year = "2020"
}

@article{Adamek:2014gva,
    author = "Adamek, Julian and Clarkson, Chris and Durrer, Ruth and Kunz, Martin",
    title = "{Does small scale structure significantly affect cosmological dynamics?}",
    eprint = "1408.2741",
    archivePrefix = "arXiv",
    primaryClass = "astro-ph.CO",
    doi = "10.1103/PhysRevLett.114.051302",
    journal = "Phys. Rev. Lett.",
    volume = "114",
    number = "5",
    pages = "051302",
    year = "2015"
}

@article{Maartens:2011yx,
    author = "Maartens, Roy",
    title = "{Is the Universe homogeneous?}",
    eprint = "1104.1300",
    archivePrefix = "arXiv",
    primaryClass = "astro-ph.CO",
    doi = "10.1098/rsta.2011.0289",
    journal = "Phil. Trans. Roy. Soc. Lond. A",
    volume = "369",
    pages = "5115--5137",
    year = "2011"
}

@article{Fleury:2025ykl,
    author = "Fleury, Pierre",
    title = "{Elements of cosmology beyond FLRW}",
    eprint = "2509.01320",
    archivePrefix = "arXiv",
    primaryClass = "astro-ph.CO",
    doi = "10.21468/SciPostPhysLectNotes.125",
    journal = "SciPost Phys. Lect. Notes",
    volume = "125",
    pages = "1",
    year = "2026"
}

@article{Ellis:1987zz,
    author = "Ellis, G. F. R. and Stoeger, W.",
    title = "{The 'fitting problem' in cosmology}",
    doi = "10.1088/0264-9381/4/6/025",
    journal = "Class. Quant. Grav.",
    volume = "4",
    pages = "1697--1729",
    year = "1987"
}

@article{Ellis:1984bqf,
    author = "Ellis, G. F. R.",
    editor = "Bertotti, B. and de Felice, F. and Pascolini, A.",
    title = "{Relativistic Cosmology: Its Nature, Aims and Problems}",
    doi = "10.1007/978-94-009-6469-3_14",
    journal = "Fundam. Theor. Phys.",
    volume = "9",
    pages = "215--288",
    year = "1984"
}

@article{Buchert:2019mvq,
    author = "Buchert, Thomas and Mourier, Pierre and Roy, Xavier",
    title = "{On average properties of inhomogeneous fluids in general relativity III: general fluid cosmologies}",
    eprint = "1912.04213",
    archivePrefix = "arXiv",
    primaryClass = "gr-qc",
    doi = "10.1007/s10714-020-02670-6",
    journal = "Gen. Rel. Grav.",
    volume = "52",
    number = "3",
    pages = "27",
    year = "2020"
}

@article{Buchert:2018yhd,
    author = "Buchert, Thomas and Mourier, Pierre and Roy, Xavier",
    title = "{Cosmological backreaction and its dependence on spacetime foliation}",
    eprint = "1805.10455",
    archivePrefix = "arXiv",
    primaryClass = "gr-qc",
    doi = "10.1088/1361-6382/aaebce",
    journal = "Class. Quant. Grav.",
    volume = "35",
    number = "24",
    pages = "24LT02",
    year = "2018"
}

@article{Buchert:2006ya,
    author = "Buchert, Thomas and Larena, Julien and Alimi, Jean-Michel",
    title = "{Correspondence between kinematical backreaction and scalar field cosmologies: The `Morphon field'}",
    eprint = "gr-qc/0606020",
    archivePrefix = "arXiv",
    doi = "10.1088/0264-9381/23/22/018",
    journal = "Class. Quant. Grav.",
    volume = "23",
    pages = "6379--6408",
    year = "2006"
}

@article{Hu:2024qnx,
    author = "Hu, Jianping and Hu, Jian and Jia, Xuandong and Gao, Baoquan and Wang, Fayin",
    title = "{Testing cosmic anisotropy with Pad{\'e} approximations and the latest Pantheon+ sample}",
    eprint = "2406.14827",
    archivePrefix = "arXiv",
    primaryClass = "astro-ph.CO",
    doi = "10.1051/0004-6361/202450342",
    journal = "Astron. Astrophys.",
    volume = "689",
    pages = "A215",
    year = "2024"
}

@article{Bengaly:2024ree,
    author = "Bengaly, Carlos A. P. and Alcaniz, Jailson S. and Pigozzo, C{\'a}ssio",
    title = "{Testing the isotropy of cosmic acceleration with the Pantheon+ and SH0ES datasets: A cosmographic analysis}",
    eprint = "2402.17741",
    archivePrefix = "arXiv",
    primaryClass = "astro-ph.CO",
    doi = "10.1103/PhysRevD.109.123533",
    journal = "Phys. Rev. D",
    volume = "109",
    number = "12",
    pages = "123533",
    year = "2024"
}

@article{Tang:2023kzs,
    author = "Tang, Li and Lin, Hai-Nan and Liu, Liang and Li, Xin",
    title = "{Consistency of Pantheon+ supernovae with a large-scale isotropic universe*}",
    eprint = "2309.11320",
    archivePrefix = "arXiv",
    primaryClass = "astro-ph.CO",
    reportNumber = "Chinese Physics C Vol. 47, No. 12 (2023) 125101",
    doi = "10.1088/1674-1137/acfaf0",
    journal = "Chin. Phys. C",
    volume = "47",
    number = "12",
    pages = "125101",
    year = "2023"
}

@article{Secrest:2022uvx,
    author = "Secrest, Nathan J. and von Hausegger, Sebastian and Rameez, Mohamed and Mohayaee, Roya and Sarkar, Subir",
    title = "{A Challenge to the Standard Cosmological Model}",
    eprint = "2206.05624",
    archivePrefix = "arXiv",
    primaryClass = "astro-ph.CO",
    doi = "10.3847/2041-8213/ac88c0",
    journal = "Astrophys. J. Lett.",
    volume = "937",
    number = "2",
    pages = "L31",
    year = "2022"
}

@article{Dam:2022wwh,
    author = "Dam, Lawrence and Lewis, Geraint F. and Brewer, Brendon J.",
    title = "{Testing the cosmological principle with CatWISE quasars: a bayesian analysis of the number-count dipole}",
    eprint = "2212.07733",
    archivePrefix = "arXiv",
    primaryClass = "astro-ph.CO",
    doi = "10.1093/mnras/stad2322",
    journal = "Mon. Not. Roy. Astron. Soc.",
    volume = "525",
    number = "1",
    pages = "231--245",
    year = "2023"
}

@article{Aluri:2022hzs,
    author = "Aluri, Pavan Kumar and others",
    title = "{Is the observable Universe consistent with the cosmological principle?}",
    eprint = "2207.05765",
    archivePrefix = "arXiv",
    primaryClass = "astro-ph.CO",
    doi = "10.1088/1361-6382/acbefc",
    journal = "Class. Quant. Grav.",
    volume = "40",
    number = "9",
    pages = "094001",
    year = "2023"
}

@article{Clarkson:2010uz,
    author = "Clarkson, Chris and Maartens, Roy",
    title = "{Inhomogeneity and the foundations of concordance cosmology}",
    eprint = "1005.2165",
    archivePrefix = "arXiv",
    primaryClass = "astro-ph.CO",
    doi = "10.1088/0264-9381/27/12/124008",
    journal = "Class. Quant. Grav.",
    volume = "27",
    pages = "124008",
    year = "2010"
}

@article{Fleury:2013sna,
    author = "Fleury, Pierre and Dupuy, H{\'e}l{\`e}ne and Uzan, Jean-Philippe",
    title = "{Interpretation of the Hubble diagram in a nonhomogeneous universe}",
    eprint = "1302.5308",
    archivePrefix = "arXiv",
    primaryClass = "astro-ph.CO",
    doi = "10.1103/PhysRevD.87.123526",
    journal = "Phys. Rev. D",
    volume = "87",
    number = "12",
    pages = "123526",
    year = "2013"
}

@article{Adamek:2017mzb,
    author = "Adamek, Julian and Clarkson, Chris and Daverio, David and Durrer, Ruth and Kunz, Martin",
    title = "{Safely smoothing spacetime: backreaction in relativistic cosmological simulations}",
    eprint = "1706.09309",
    archivePrefix = "arXiv",
    primaryClass = "astro-ph.CO",
    doi = "10.1088/1361-6382/aaeca5",
    journal = "Class. Quant. Grav.",
    volume = "36",
    number = "1",
    pages = "014001",
    year = "2019"
}

@article{Wiltshire:2011vy,
    author = "Wiltshire, David L.",
    title = "{What is dust? - Physical foundations of the averaging problem in cosmology}",
    eprint = "1106.1693",
    archivePrefix = "arXiv",
    primaryClass = "gr-qc",
    doi = "10.1088/0264-9381/28/16/164006",
    journal = "Class. Quant. Grav.",
    volume = "28",
    pages = "164006",
    year = "2011"
}

@article{Perivolaropoulos:2014lua,
    author = "Perivolaropoulos, Leandros",
    title = "{Large Scale Cosmological Anomalies and Inhomogeneous Dark Energy}",
    eprint = "1401.5044",
    archivePrefix = "arXiv",
    primaryClass = "astro-ph.CO",
    doi = "10.3390/galaxies2010022",
    journal = "Galaxies",
    volume = "2",
    pages = "22--61",
    year = "2014"
}

@article{Zumalacarregui:2012pq,
    author = "Zumalacarregui, Miguel and Garcia-Bellido, Juan and Ruiz-Lapuente, Pilar",
    title = "{Tension in the Void: Cosmic Rulers Strain Inhomogeneous Cosmologies}",
    eprint = "1201.2790",
    archivePrefix = "arXiv",
    primaryClass = "astro-ph.CO",
    reportNumber = "IFT-UAM-CSIC-12-01",
    doi = "10.1088/1475-7516/2012/10/009",
    journal = "JCAP",
    volume = "10",
    pages = "009",
    year = "2012"
}

@article{Bull:2011wi,
    author = "Bull, Philip and Clifton, Timothy and Ferreira, Pedro G.",
    title = "{The kSZ effect as a test of general radial inhomogeneity in LTB cosmology}",
    eprint = "1108.2222",
    archivePrefix = "arXiv",
    primaryClass = "astro-ph.CO",
    doi = "10.1103/PhysRevD.85.024002",
    journal = "Phys. Rev. D",
    volume = "85",
    pages = "024002",
    year = "2012"
}

@article{EGS1968,
  author  = {Ehlers, J. and Geren, P. and Sachs, R. K.},
  title   = {Isotropic solutions of the {Einstein--Liouville} equations},
  journal = {J. Math. Phys.},
  volume  = {9},
  pages   = {1344--1349},
  year    = {1968},
  doi     = {10.1063/1.1664720}
}

@article{Verlinde:2016toy,
    author = "Verlinde, Erik P.",
    title = "{Emergent Gravity and the Dark Universe}",
    eprint = "1611.02269",
    archivePrefix = "arXiv",
    primaryClass = "hep-th",
    doi = "10.21468/SciPostPhys.2.3.016",
    journal = "SciPost Phys.",
    volume = "2",
    number = "3",
    pages = "016",
    year = "2017"
}

@article{Tuveri:2019zor,
    author = "Tuveri, Matteo and Cadoni, Mariano",
    title = "{Galactic dynamics and long-range quantum gravity}",
    eprint = "1904.11835",
    archivePrefix = "arXiv",
    primaryClass = "gr-qc",
    doi = "10.1103/PhysRevD.100.024029",
    journal = "Phys. Rev. D",
    volume = "100",
    number = "2",
    pages = "024029",
    year = "2019"
}

@article{DESI2024VI,
    author = "Adame, A. G. and others",
    collaboration = "DESI",
    title = "{DESI 2024 VI: cosmological constraints from the measurements of baryon acoustic oscillations}",
    eprint = "2404.03002",
    archivePrefix = "arXiv",
    primaryClass = "astro-ph.CO",
    reportNumber = "FERMILAB-PUB-24-0154-PPD",
    doi = "10.1088/1475-7516/2025/02/021",
    journal = "JCAP",
    volume = "02",
    pages = "021",
    year = "2025"
}

@article{DESIDR2II,
    author = "Abdul Karim, M. and others",
    collaboration = "DESI",
    title = "{DESI DR2 results. II. Measurements of baryon acoustic oscillations and cosmological constraints}",
    eprint = "2503.14738",
    archivePrefix = "arXiv",
    primaryClass = "astro-ph.CO",
    reportNumber = "FERMILAB-PUB-25-0169-PPD",
    doi = "10.1103/tr6y-kpc6",
    journal = "Phys. Rev. D",
    volume = "112",
    number = "8",
    pages = "083515",
    year = "2025"
}

@article{BuchertRasanen2012,
    author = {Buchert, Thomas and R{\"a}s{\"a}nen, Syksy},
    title = "{Backreaction in late-time cosmology}",
    eprint = "1112.5335",
    archivePrefix = "arXiv",
    primaryClass = "astro-ph.CO",
    reportNumber = "HIP-2011-35-TH",
    doi = "10.1146/annurev.nucl.012809.104435",
    journal = "Ann. Rev. Nucl. Part. Sci.",
    volume = "62",
    pages = "57--79",
    year = "2012"
}

@article{Wiltshire2007b,
   author = "Wiltshire, David L.",
    title = "{Exact solution to the averaging problem in cosmology}",
    eprint = "0709.0732",
    archivePrefix = "arXiv",
    primaryClass = "gr-qc",
    doi = "10.1103/PhysRevLett.99.251101",
    journal = "Phys. Rev. Lett.",
    volume = "99",
    pages = "251101",
    year = "2007"
}

@article{Buchert2000,
    author = "Buchert, Thomas",
    title = "{On average properties of inhomogeneous fluids in general relativity. 1. Dust cosmologies}",
    eprint = "gr-qc/9906015",
    archivePrefix = "arXiv",
    reportNumber = "CERN-TH-99-165",
    doi = "10.1023/A:1001800617177",
    journal = "Gen. Rel. Grav.",
    volume = "32",
    pages = "105--125",
    year = "2000"
}

@article{Buchert2001,
    author = "Buchert, Thomas",
    title = "{On average properties of inhomogeneous fluids in general relativity: Perfect fluid cosmologies}",
    eprint = "gr-qc/0102049",
    archivePrefix = "arXiv",
    doi = "10.1023/A:1012061725841",
    journal = "Gen. Rel. Grav.",
    volume = "33",
    pages = "1381--1405",
    year = "2001"
}

@article{Cadoni:2018dnd,
    author = "Cadoni, M. and Casadio, R. and Giusti, A. and Tuveri, M.",
    title = "{Emergence of a Dark Force in Corpuscular Gravity}",
    eprint = "1801.10374",
    archivePrefix = "arXiv",
    primaryClass = "gr-qc",
    doi = "10.1103/PhysRevD.97.044047",
    journal = "Phys. Rev. D",
    volume = "97",
    number = "4",
    pages = "044047",
    year = "2018"
}

@article{MarraNotari2011,
  author = "Marra, Valerio and Notari, Alessio",
    title = "{Observational constraints on inhomogeneous cosmological models without dark energy}",
    eprint = "1102.1015",
    archivePrefix = "arXiv",
    primaryClass = "astro-ph.CO",
    doi = "10.1088/0264-9381/28/16/164004",
    journal = "Class. Quant. Grav.",
    volume = "28",
    pages = "164004",
    year = "2011"
}

@article{GarciaBellidoHaugboelle2008,
    author = "Garcia-Bellido, Juan and Haugboelle, Troels",
    title = "{Confronting Lemaitre-Tolman-Bondi models with Observational Cosmology}",
    eprint = "0802.1523",
    archivePrefix = "arXiv",
    primaryClass = "astro-ph",
    reportNumber = "IFT-UAM-CSIC-08-03",
    doi = "10.1088/1475-7516/2008/04/003",
    journal = "JCAP",
    volume = "04",
    pages = "003",
    year = "2008"
}

@article{Szekeres1975,
  author  = {Szekeres, P.},
  title   = {A class of inhomogeneous cosmological models},
  journal = {Commun. Math. Phys.},
  volume  = {41},
  pages   = {55--64},
  year    = {1975},
  doi     = {10.1007/BF01608547}
}

@article{Bondi1947,
  author  = {Bondi, H.},
  title   = {Spherically symmetrical models in general relativity},
  journal = {Mon. Not. R. Astron. Soc.},
  volume  = {107},
  pages   = {410--425},
  year    = {1947},
  doi     = {10.1093/mnras/107.5-6.410}
}

@article{Tolman:1934za,
    author = "Tolman, Richard C.",
    title = "{Effect of imhomogeneity on cosmological models}",
    doi = "10.1073/pnas.20.3.169",
    journal = "Proc. Nat. Acad. Sci.",
    volume = "20",
    pages = "169--176",
    year = "1934"
}

@article{Lemaitre:1933gd,
    author = "Lemaitre, G.",
    title = "{The expanding universe}",
    doi = "10.1023/A:1018855621348",
    journal = "Annales Soc. Sci. Bruxelles A",
    volume = "53",
    pages = "51--85",
    year = "1933"
}

@article{Bolejko:2011jc,
    author = "Bolejko, Krzysztof and Celerier, Marie-Noelle and Krasinski, Andrzej",
    title = "{Inhomogeneous cosmological models: Exact solutions and their applications}",
    eprint = "1102.1449",
    archivePrefix = "arXiv",
    primaryClass = "astro-ph.CO",
    doi = "10.1088/0264-9381/28/16/164002",
    journal = "Class. Quant. Grav.",
    volume = "28",
    pages = "164002",
    year = "2011"
}

@article{Clarkson:2011zq,
    author = "Clarkson, Chris and Ellis, George and Larena, Julien and Umeh, Obinna",
    title = "{Does the growth of structure affect our dynamical models of the universe? The averaging, backreaction and fitting problems in cosmology}",
    eprint = "1109.2314",
    archivePrefix = "arXiv",
    primaryClass = "astro-ph.CO",
    doi = "10.1088/0034-4885/74/11/112901",
    journal = "Rept. Prog. Phys.",
    volume = "74",
    pages = "112901",
    year = "2011"
}

@article{Sawala:2025ajg,
    author = "Sawala, Till and Teeriaho, Meri and Frenk, Carlos S. and Helly, John and Jenkins, Adrian and Racz, Gabor and Schaller, Matthieu and Schaye, Joop",
    title = "{The Emperor's New Arc: gigaparsec patterns abound in a $\Lambda$CDM universe}",
    eprint = "2502.03515",
    archivePrefix = "arXiv",
    primaryClass = "astro-ph.CO",
    month = "2",
    year = "2025"
}

@article{Lopez:2024kih,
    author = "Lopez, Alexia M. and Clowes, Roger G. and Williger, Gerard M.",
    title = "{Investigating ultra-large large-scale structures: potential implications for cosmology}",
    eprint = "2409.14894",
    archivePrefix = "arXiv",
    primaryClass = "astro-ph.CO",
    doi = "10.1098/rsta.2024.0029",
    journal = "Phil. Trans. Roy. Soc. Lond. A",
    volume = "383",
    number = "2290",
    pages = "20240029",
    year = "2025"
}

@ARTICLE{2013MNRAS.429.2910C,
       author = {{Clowes}, Roger G. and {Harris}, Kathryn A. and {Raghunathan}, Srinivasan and {Campusano}, Luis E. and {S{\"o}chting}, Ilona K. and {Graham}, Matthew J.},
        title = "{A structure in the early Universe at z {\ensuremath{\sim}} 1.3 that exceeds the homogeneity scale of the R-W concordance cosmology}",
     journal = {Mon. Not. Roy. Astron. Soc.},
         year = 2013,
        month = mar,
       volume = {429},
       number = {4},
        pages = {2910-2916},
          doi = {10.1093/mnras/sts497},
archivePrefix = {arXiv},
       eprint = {1211.6256},
 primaryClass = {astro-ph.CO},
       adsurl = {https://ui.adsabs.harvard.edu/abs/2013MNRAS.429.2910C}
}

@article{Yadav:2005vv,
    author = "Yadav, Jaswant and Bharadwaj, Somnath and Pandey, Biswajit and Seshadri, T. R.",
    title = "{Testing homogeneity on large scales in the Sloan Digital Sky Survey Data Release One}",
    eprint = "astro-ph/0504315",
    archivePrefix = "arXiv",
    doi = "10.1111/j.1365-2966.2005.09578.x",
    journal = "Mon. Not. Roy. Astron. Soc.",
    volume = "364",
    pages = "601--606",
    year = "2005"
}

@book{Peebles:1980yev,
    author = "Peebles, P. James",
    title = "{The Large-Scale Structure of the Universe}",
    isbn = "978-0-691-08240-0, 978-0-691-20983-8, 978-0-691-20671-4",
    publisher = "Princeton University Press",
    month = "11",
    year = "1980"
}

@article{Herrera:1997plx,
	author = {Herrera, L. and Santos, N. O.},
	doi = {10.1016/S0370-1573(96)00042-7},
	journal = {Phys. Rept.},
	pages = {53--130},
	title = {{Local anisotropy in self-gravitating systems}},
	volume = {286},
	year = {1997}}

@article{Ellis:1998ct,
	archiveprefix = {arXiv},
	author = {Ellis, George F. R. and van Elst, Henk},
	doi = {10.1007/978-94-011-4455-1_1},
	eprint = {gr-qc/9812046},
	journal = {NATO Sci. Ser. C},
	pages = {1--116},
	title = {{Cosmological models: Cargese lectures 1998}},
	volume = {541},
	year = {1999}}

@article{Horvat:2010xf,
	archiveprefix = {arXiv},
	author = {Horvat, Dubravko and Ilijic, Sasa and Marunovic, Anja},
	doi = {10.1088/0264-9381/28/2/025009},
	eprint = {1010.0878},
	journal = {Class. Quant. Grav.},
	pages = {025009},
	primaryclass = {gr-qc},
	title = {{Radial pulsations and stability of anisotropic stars with quasi-local equation of state}},
	volume = {28},
	year = {2011}}

@article{Misner:1964je,
    author = "Misner, Charles W. and Sharp, David H.",
    title = "{Relativistic equations for adiabatic, spherically symmetric gravitational collapse}",
    doi = "10.1103/PhysRev.136.B571",
    journal = "Phys. Rev.",
    volume = "136",
    pages = "B571--B576",
    year = "1964"
}

@article{Cadoni:2024rri,
    author = "Cadoni, Mariano and Pitzalis, Mirko and Sanna, Andrea Pierfrancesco",
    title = "{Apparent horizons in cosmologically-embedded black holes}",
    eprint = "2410.10459",
    archivePrefix = "arXiv",
    primaryClass = "gr-qc",
    doi = "10.1088/1475-7516/2025/02/051",
    journal = "JCAP",
    volume = "02",
    pages = "051",
    year = "2025"
}
\end{document}